\newcommand{\benchmark}{ComUIBench}
\newcommand{\scheme}{ComUICoder}
\documentclass[sigconf]{acmart}
\usepackage{algorithmic}
\usepackage[ruled, linesnumbered]{algorithm2e}
\usepackage{subfigure}
\usepackage{multirow}
\usepackage{enumerate}
\usepackage{pifont}

\usepackage{amsmath}

\usepackage{listings}
\usepackage{xcolor}
\usepackage{colortbl}
\usepackage{tikz}
\usepackage{tcolorbox}
\tcbuselibrary{skins,breakable}
\usepackage{enumitem}

\newcommand{\yes}{\color{green!60!black}\ding{51}}
\newcommand{\no}{\color{red!60!black}\ding{55}} 
\usepackage{titletoc}

\renewcommand{\arraystretch}{0.8}

\acmConference[Conference ’26]{Make sure to enter the correct
  conference title from your rights confirmation emai}{July, 
  2026}{DC, USA}

\begin{document}






\title{ComUICoder: Component-based Reusable UI Code Generation \\ for Complex Websites via Semantic Segmentation and Element-wise Feedback}




\author{Jingyu Xiao}
\affiliation{%
  \institution{The Chinese University of Hong Kong}
  \city{Hong Kong}
  \country{China}}
\email{jyxiao@link.cuhk.edu.hk}

\author{Jiantong Qin}
\affiliation{%
  \institution{The Chinese University of Hong Kong}
  \city{Hong Kong}
  \country{China}}
\email{jtqin@link.cuhk.edu.hk}

\author{Shuoqi Li}
\affiliation{%
  \institution{The Chinese University of Hong Kong}
  \city{Hong Kong}
  \country{China}}
\email{1155191595@link.cuhk.edu.hk}

\author{Man Ho LAM}
\affiliation{%
  \institution{The Chinese University of Hong Kong}
  \city{Hong Kong}
  \country{China}}
\email{mhlam@link.cuhk.edu.hk}

\author{Yuxuan Wan}
\affiliation{%
  \institution{The Chinese University of Hong Kong}
  \city{Hong Kong}
  \country{China}}
\email{yxwan@link.cuhk.edu.hk}

\author{Jen-tse Huang}
\affiliation{%
  \institution{Johns Hopkins University}
  \city{Baltimore}
  \country{USA}}
\email{jhuan236@jh.edu}

\author{Yintong Huo}
\affiliation{%
  \institution{Singapore Management University}
  \country{Singapore}}
\email{ythuo@smu.edu.sg}

\author{Michael R. Lyu}
\affiliation{%
  \institution{The Chinese University of Hong Kong}
  \city{Hong Kong}
  \country{China}}
\email{lyu@cse.cuhk.edu.hk}



\begin{abstract}
Multimodal Large Language Models (MLLMs) have demonstrated strong performance on the UI-to-code task, which aims to generate UI code from design mock-ups. However, when applied to long and complex websites, they often struggle with fragmented segmentation, redundant code generation for repetitive components, and frequent UI inconsistencies. To systematically investigate and address these challenges, we introduce \benchmark, a new multi-page complex webpage benchmark with component annotations, designed to evaluate MLLMs’ ability to generate reusable UI code in realistic website scenarios. Building upon this benchmark, we propose \scheme, a component-based UI code generation framework that emphasizes semantic-aware segmentation, code reuse, and fine-grained refinement. Specifically, \scheme \ incorporates (1) Hybrid Semantic-aware Block Segmentation for accurate UI semantic coherent block detection, (2) Visual-aware Graph-based Block Merge to consolidate structurally similar components within and across webpages for reusable implementation, and (3) Priority-based Element-wise Feedback to refine generated code and reduce element-level inconsistencies. Extensive experiments demonstrate that \scheme \ significantly improves overall generation quality and code reusability on complex multi-page websites. Our datasets and code are publicly available at \url{https://github.com/WebPAI/ComUICoder}.

\end{abstract}






\maketitle

\section{Introduction}

Converting webpage designs into functional UI code is a critical yet labor-intensive step in website development. Owing to their strong visual understanding capabilities~\cite{liu2025benchmarking} and code generation abilities~\cite{tang2025slidecoder, gao2025treat}, Multimodal Large Language Models (MLLMs) have been widely adopted for UI-to-Code tasks~\cite{wan2025divide, wan2024mrweb, wan2025automatically, xiao2024interaction2code, xiao2025designbench, xiao2025efficientuicoder, dang2025envisioning}, which aim to generate code that faithfully reproduces webpage elements, layouts, text, and color schemes from UI designs.


Despite promising results reported by existing methods~\cite{wan2025divide, gui2025latcoder, wu2025mllm}, real-world commercial websites typically involve large-scale, multi-page, and feature-rich webpages. When applied to such realistic environments, MLLMs commonly face the following challenges:



\begin{figure}[ht]
    \centering
    \includegraphics[width = .45\textwidth]{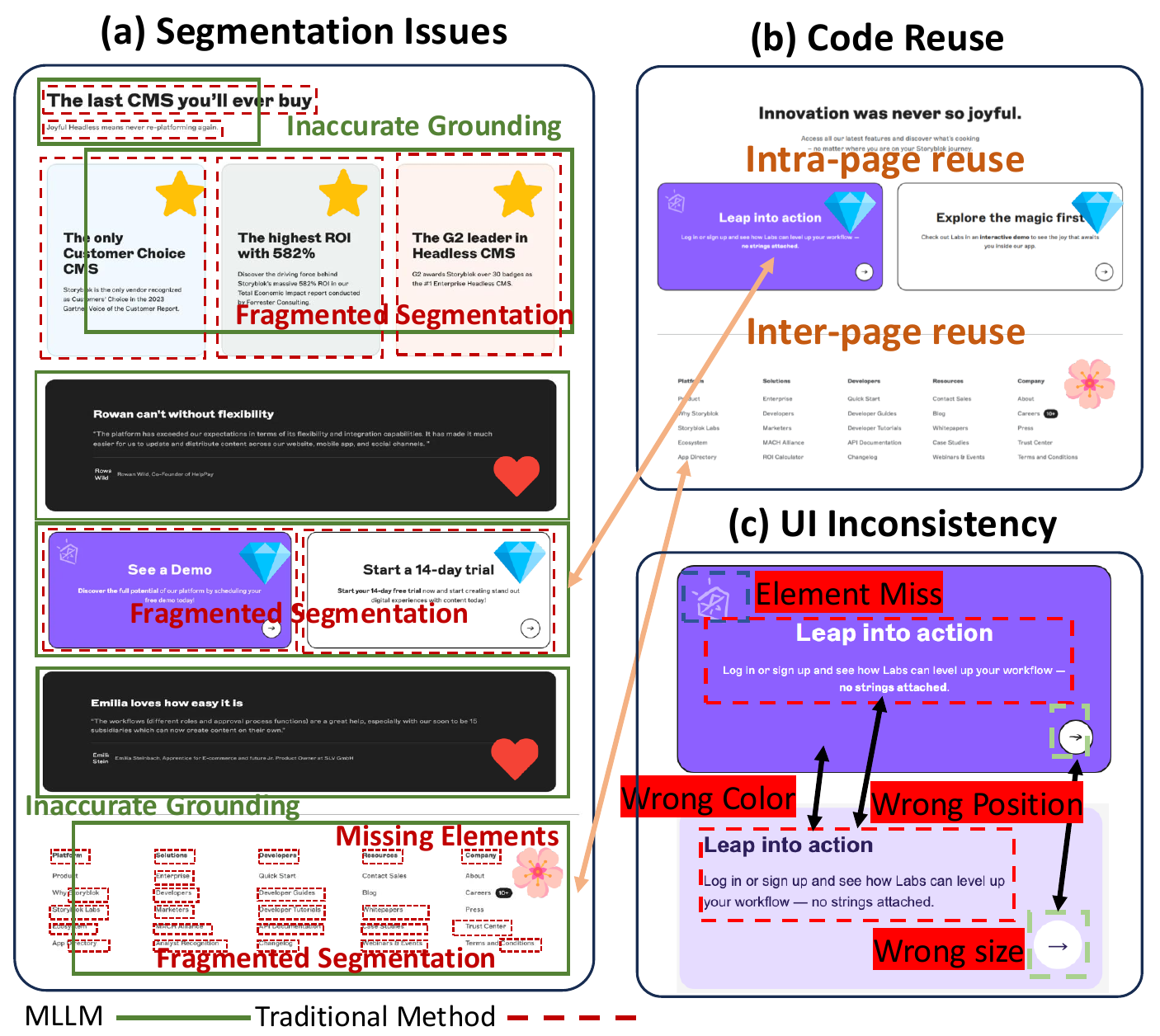}
    \caption{Problems encountered by the model when facing complex websites. The same icon denotes the components with similar structure. The bounding box denotes the segmentation results.}
    \label{fig:example}
\end{figure}

\begin{table}[]
\caption{Comparisons between  \benchmark/\scheme \ and existing UI2Code benchmarks/methods.}
\footnotesize
\resizebox{\linewidth}{!}{
\setlength{\tabcolsep}{.12em}{
\begin{tabular}{@{}l|c|c|c@{}}
\toprule
\multicolumn{4}{c}{Benchmarks}                                       \\ \midrule
            & Complexity        & Multi-page  & \multicolumn{1}{c}{\begin{tabular}[c]{@{}c@{}}Component\\ Annotation\end{tabular}}  \\ \midrule
WebSight~\cite{laurençon2024unlocking}    & Simple                & \no         & \no                    \\
Design2Code~\cite{si2024design2code} & Simple                & \no         & \no                    \\
Web2Code~\cite{yun2024web2code}    & Simple                & \no         & \no                    \\
WebCode2M~\cite{gui2025webcode2m}   & Simple                & \no         & \no    \\
DesignBench~\cite{xiao2025designbench} & Medium                & \no         & \no   \\
ComUIBench (Ours)  & Complex               & \yes        & \yes                   \\ \toprule
\multicolumn{4}{c}{Methods}                                          \\ \midrule
            & {\begin{tabular}[c]{@{}c@{}}Semantic-aware\\ Segmentation\end{tabular}} & Code Reuse & {\begin{tabular}[c]{@{}c@{}}Fine-grained\\ Feedback\end{tabular}} \\ \midrule

DCGen~\cite{wan2025divide}       & \no                & \no         & \no                    \\
UICopilot~\cite{gui2025uicopilot} & \no                & \no         & \no                    \\
LayoutCoder~\cite{wu2025mllm} & \no                & \no         & \no                    \\
LatCoder~\cite{gui2025latcoder}    & \no                & \no         & \no                    \\
ComUICoder (Ours)  & \yes               & \yes        & \yes                    \\ \bottomrule
\end{tabular}}}
\end{table}


\textbf{MLLMs struggle with intricate layouts and excessively long webpages.} Existing methods adopt either model-based~\cite{gui2025uicopilot} or rule-based~\cite{wan2025divide, gui2025latcoder} segmentation within a segment-and-assemble pipeline. Model-based approaches suffer from inaccurate grounding due to limited MLLM capabilities (green box in Fig.~\ref{fig:example}(a)), while rule-based methods, though precise in boundary detection, often over-segment webpages, disrupting structural and semantic coherence (red boxes in Fig.~\ref{fig:example}(a)). This over-segmentation leads to higher generation costs and increased assembly complexity.

\textbf{MLLMs tend to generate duplicate code for similar components, reducing code reusability and maintainability.} Webpages frequently contain components with similar hierarchical layouts. As illustrated in Figure~\ref{fig:example} (a) and (b), such components may appear adjacently (e.g., those marked with star \ding{72} and heart \ding{170} icons), be spatially separated on the same page (e.g., those marked with diamond $\blacktriangledown$ icons), or recur across different webpages of the same site (e.g., those marked with flower \ding{96} icons). When processing these repetitive components, MLLMs typically generate redundant code for each instance rather than abstracting them into reusable components, which significantly undermines code maintainability.

\textbf{MLLMs fail to precisely preserve visual and layout fidelity when reproducing webpage.} The rendered UI may deviate from the original design due to the limited ability of MLLMs to accurately interpret webpages containing numerous elements and complex visual details. As shown in Figure~\ref{fig:example}(c), such inconsistencies include mismatches in size, color, position, element missing and so on.

To address the challenges, we construct \benchmark, a multi-page complex website benchmark, which is significantly more complex than existing UI2Code benchmarks with multiple subpages and component annotations, to evaluate MLLMs' ability to generate complex webpages. We then propose \scheme, a component-based UI code generation framework. \scheme \ comprises three core innovations: First, \textbf{Hybrid Semantic-aware Block Segmentation (HSBS)} combines MLLMs' semantic reasoning with UI detection tools to achieve both semantically coherent and spatially precise UI block detection. Second, \textbf{Visual-aware Graph-based Block Merge (VGBM)} employs graph-based structural matching and visual similarity metrics to cluster reusable components across pages, enabling component-based implementation. Third, \textbf{Priority-based Element-wise Feedback (PEF)} performs element-level comparison between generated and reference UIs, generating targeted refinement instructions prioritized by visual saliency and error severity. Our contributions are summarized as following:










\begin{itemize}[leftmargin=*]
    

    \item \textbf{Benchmarks}. We build the first multi-page complex UI2Code benchmark \textbf{\benchmark} \ containing 40 web projects with 150 webpages and 2,055 annotated UI semantic coherent blocks  grouped into 1,134 component groups to evaluate MLLM's ability to implement complex webpages.

    \item \textbf{Methods.} We propose ComUICoder, a component-based framework featuring: (1) HSBS for accurate UI semantic coherent block detection; (2) VGBM for intra- and inter-page component reuse via visual-aware graph-based matching; and (3) PEF for fine-grained element-wise UI inconsistencies refinement.
    

    \item \textbf{Evaluation.} Extensive experiments show \scheme \ significantly outperforms SOTA methods in visual fidelity and code reusability.


\end{itemize}




\section{Background}



\subsection{Task Definition}

UI2Code aims to convert a visual webpage mockup into functional HTML+CSS code that faithfully reproduces its layout and appearance. Given a UI mockup~\cite{uimockup} $I_0$, a model $M$ generates code $C_g = M(I_0)$, whose rendered output $I_g$ should closely match $I_0$ in structure and visual fidelity.

\subsection{Related Work}
\textbf{UI2Code Methods.} Existing methods follow three paradigms: (1) Deep learning-based approaches learn direct image-to-code mappings through CNNs~\cite{acsirouglu2019automatic, cizotto2023web, moran2018machine, Xu2021Image2e, Chen2018FromUI}, CNN-LSTM architectures~\cite{beltramelli2018pix2code}, and attention-based encoder-decoders~\cite{chen2022code}. (2) Computer vision-based methods emphasize structural understanding via object detection~\cite{jain2019sketch2code} and OCR~\cite{nguyen2015reverse}. (3) MLLM-based approaches address complex webpage challenges through divide-and-conquer strategies~\cite{wan2025divide}, layout-aware modeling~\cite{wu2025mllm, gui2025latcoder}, and token compression~\cite{xiao2025efficientuicoder}. \textit{However, these methods lack semantic-aware segmentation, code reusability, and fine-grained feedback mechanisms.} \textbf{UI2Code Benchmarks.} Benchmarks have evolved from synthetic datasets like WebSight and Web2Code~\cite{yun2024web2code} to real-world collections including Design2Code~\cite{si2024design2code} (484 curated webpages) and WebCode2M~\cite{gui2025webcode2m} (20,000 samples). Task-specific benchmarks target interactive UIs~\cite{xiao2024interaction2code}, multi-page websites~\cite{wan2024mrweb}, and framework-based generation~\cite{xiao2025designbench}. \textit{Despite this progress, existing benchmarks focus on simple webpages and lack component-level annotations and code reusability evaluation.}

\section{The \benchmark \ Benchmark}




\subsection{Dataset Collection}

\benchmark\ is built through a carefully designed data collection and annotation pipeline. We recruit five PhD students with professional front-end development experience to curate representative websites from the Moz Top 500 rankings~\cite{moz_top500}, ensuring high-impact and real-world design patterns. Website selection follows comprehensive criteria emphasizing design complexity, component reusability across pages, structural diversity with multiple page templates, and broad domain coverage. For each website, annotators select 2–5 representative subpages to capture the core component ecosystem while avoiding redundant templates. Semantic coherent UI blocks are annotated as self-contained and reusable units with clear visual boundaries and functional integrity, following strict guidelines on boundary completeness, grouping of consecutive repeated components, and outermost-only annotation for nested structures. Finally, annotated UI blocks are grouped across pages based on structural, visual, and functional equivalence to identify reusable design patterns, enabling systematic evaluation of reusable UI code generation on complex multi-page websites. The \benchmark \ construction details are shown in Appendix~\ref{subsec:data_collection_detail}


\subsection{Data Statistics and Complexity}

The resulting \benchmark\ includes 40 websites across diverse domains, totaling 150 subpages and 2,055 annotated UI semantic coherent blocks grouped into 1,134 component groups. On average, each website has 4 subpages and 29 component groups, indicating substantial design complexity. Notably, 63.3\% of components belong to reusable groups, with 9.93 groups on average containing multiple instances per website. Detailed statistics are provided in Appendix~\ref{appendix:data_stat}. As shown in Table~\ref{tab:bench_comparison}, \benchmark\ exhibits the highest complexity, with greater DOM depth, tags and image tokens, making it more challenging to layout and structure.


\begin{table}[h]
\centering
\caption{Comparison and statistics of benchmarks. All statistics are in the "average $\pm$ standard deviation" format.}
\label{tab:bench_comparison}
\resizebox{\linewidth}{!}{
\setlength{\tabcolsep}{.1em}{
\begin{tabular}{l|cccccc}
\toprule
Benchmarks & Source & Size & {\begin{tabular}[c]{@{}c@{}}Image\\ Tokens\end{tabular}} & Tags & {\begin{tabular}[c]{@{}c@{}}DOM\\ Depth\end{tabular}} & {\begin{tabular}[c]{@{}c@{}}Unique\\ Tags\end{tabular}} \\
\midrule
WebSight~\cite{laurençon2024unlocking} & Synthetic & 0.8M & $21346 \pm 7127$ & $19 \pm 8$ & $5 \pm 1$ & $10 \pm 3$ \\
Design2Code~\cite{si2024design2code} & Real-World & 484 & $9676 \pm 4279$  & $158 \pm 100$ & $13 \pm 5$ & $22 \pm 6$ \\
Web2Code~\cite{yun2024web2code} & Real-World & 20k & $5950 \pm 2756$ & $175 \pm 94$ & $15 \pm 5$ & $21 \pm 5$ \\
WebCode2M-Test~\cite{gui2025webcode2m} & Real-World & 768 & $9557 \pm 3354$ & $149 \pm 84$ & $13 \pm 4$ & $22 \pm 6$ \\
DesignBench~\cite{xiao2025designbench} & Real-World & 430 & $17310 \pm 13414$ & $492 \pm 642$ & $17 \pm 6$ & $31 \pm 12$ \\
\benchmark \ (Ours) & Real-World & 150 & $86881\pm19771$ & \textbf{$1287 \pm 693$} & \textbf{$21 \pm 6$} & \textbf{$36 \pm 7$} \\
\bottomrule
\end{tabular}}}
\end{table}

\section{The \scheme \ Framework}

\subsection{Overview}

\begin{figure*}[ht]
    \centering
    \includegraphics[width = .99\textwidth]{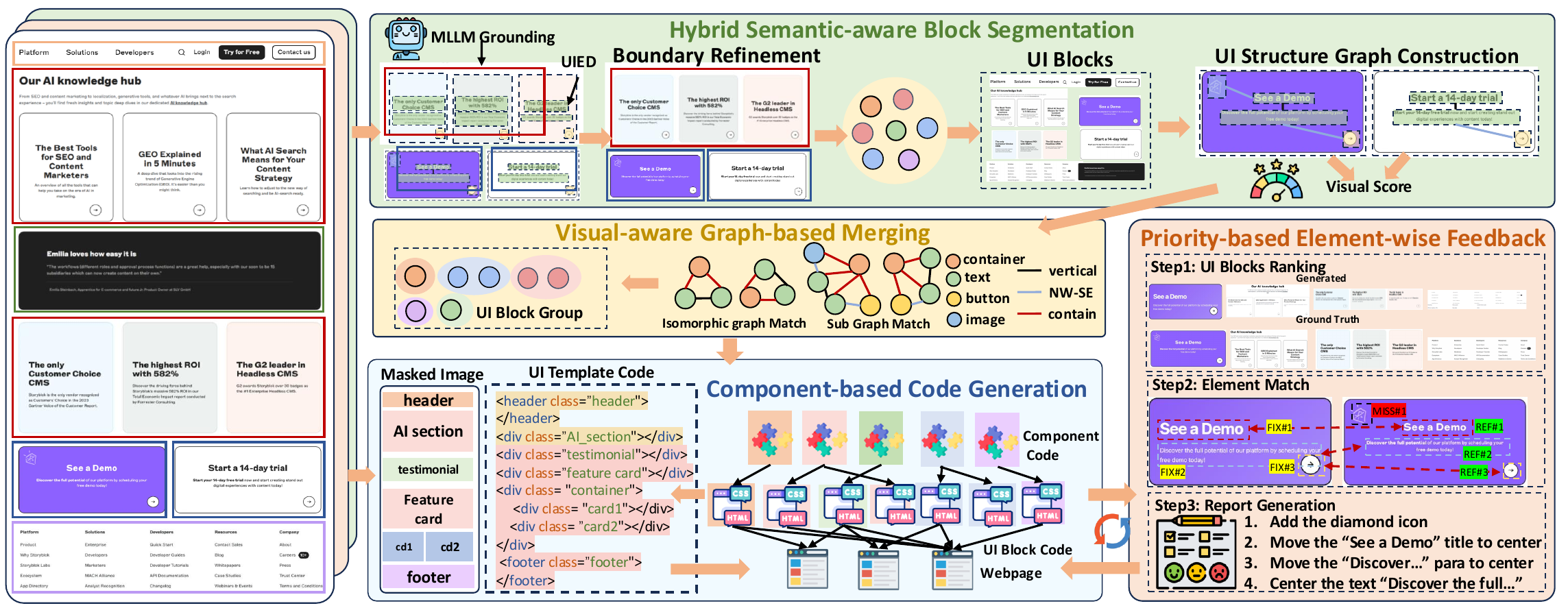}
    \caption{The overview of \scheme.}
    \label{fig:ComUICoder}
\end{figure*}

To enable accurate and reusable UI code generation for complex websites, we propose \scheme, as illustrated in Figure~\ref{fig:ComUICoder}. The workflow includes four stages: (1) Hybrid Semantic-aware Block Segmentation (HSBS) (\S\ref{subsec:HSBS}) partitions webpages into meaningful UI blocks; (2) Visual-aware Graph-based Block Merge (VGBM) (\S\ref{subsec:VGBM}) clusters visually and structurally similar blocks into reusable component groups; (3) Component-based Code Generation (CCG) (\S\ref{subsec:CCG}) generates webpage templates, individual components, and integrates them into complete webpage code; and (4) Priority-based Element-wise Feedback (PEF) (\S\ref{subsec:PEF}) refines UI inconsistencies through element-level comparison and repair.


\subsection{Hybrid Semantic-aware Block Segmentation}
\label{subsec:HSBS}



Existing segmentation methods face two major limitations: (1) they rely on rule-based approaches (e.g., divider line detection~\cite{wan2025divide} and element alignment mining~\cite{wu2025mllm}) that fail to consider the semantics of UI blocks, and (2) when applied to complex web pages, these methods~\cite{gui2025uicopilot} either fail completely or produce meaningless segmented blocks~\cite{gui2025latcoder} (e.g., blank blocks). To achieve semantic-aware and precise segmentation, we propose the Hybrid Semantic-aware Block Segmentation (HSBS) method, which combines the strengths of MLLMs and traditional UI detection tools. MLLMs excel in semantic understanding, identifying coherent UI components (e.g., navigation bars or product cards) and functional relationships, but often generate imprecise bounding boxes due to limited grounding accuracy. Conversely, tools like UIED~\cite{xie2020uied} provide precise pixel-level boundaries but lack semantic understanding, often over-segmenting components into fragmented elements (e.g., separating a button's text from its background). HSBS synergizes these approaches by leveraging MLLMs for semantic reasoning and UIED for boundary refinement, ensuring both semantic coherence and spatial accuracy.



\subsubsection{Method Formulation}

HSBS operates in a two-stage pipeline that strategically combines UIED's~\cite{xie2020uied}  precise element detection with the MLLM's semantic understanding capabilities.

\textbf{Stage 1: Traditional UI Element Detection.} We employ UIED to detect fine-grained UI elements from the webpage screenshot. Let $\mathcal{U} = \{u_1, u_2, \ldots, u_m\}$ denote the set of UI element blocks detected by UIED, where each block $u_i$ is represented by its bounding box coordinates $u_i = (x_i^{u}, y_i^{u}, w_i^{u}, h_i^{u})$. These detected elements typically correspond to low-level atomic UI components such as buttons, text fields, and images, providing pixel-accurate spatial boundaries.

\textbf{Stage 2: MLLM-based Semantic Block Detection.} We then prompt the MLLM to identify semantically meaningful UI blocks based on functional and visual coherence. To guide the MLLM's detection, we provide the following criteria for identifying semantic blocks:
(1) \textit{Block-based components}: Visual blocks (cards, panels, sections) with distinct backgrounds or borders that group related content together; (2) \textit{Blocks with associated text}: Visual blocks combined with semantically related text elements (titles, descriptions, labels) in proximity; (3) \textit{Repeated structure groups}: Collections of elements sharing similar structure and semantic purpose (e.g., product cards, image galleries, team member profiles), where each individual item is identified separately; (4) \textit{Navigational components}: Header navigation bars, footer sections, and sidebar menus that serve navigation purposes; (5) \textit{Form sections}: Related form fields grouped together for user input; (6) \textit{Content sections}: Semantically distinct content areas such as testimonials, pricing tables. The details of the prompt are shown in Appendix~\ref{subsec:prompt_detail}.


Let $\mathcal{M} = {m_1, m_2, \ldots, m_n}$ denote the semantic UI blocks identified by the MLLM, where each block $m_j = (x_j^{m}, y_j^{m}, w_j^{m}, h_j^{m})$ approximates a region of semantically coherent components.

\textbf{Stage 3: Boundary Refinement.} Given the two sets $\mathcal{U}$ and $\mathcal{M}$, we refine the MLLM-detected blocks by leveraging UIED's precise boundaries. For each MLLM block $m_j \in \mathcal{M}$, we identify the spatial relationship between $m_j$ and each UIED block $u_i \in \mathcal{U}$. We categorize these relationships into two types: {(1) Containment Relationship: When a UIED block is fully contained within an MLLM block ($u_i \subset m_j$), we retain the MLLM block's larger bounding box as it represents the complete semantic component. (2) Intersection Relationship: When a UIED block partially overlaps with an MLLM block, we compute the overlap ratio: $r_{ij} = \frac{\text{Area}(u_i \cap m_j)}{\text{Area}(u_i)}$, where $\text{Area}(\cdot)$ denotes the area of a bounding box. We merge the UIED block into the MLLM block only when $r_{ij} > t_{overlap}$, where $t_{overlap}$ is a predefined threshold. Specifically, we compute the refined intersection set: $
\mathcal{I}_j = \{u_i \in \mathcal{U} \mid (u_i \subset m_j) \lor (r_{ij} > t_{overlap} \land u_i \cap m_j \neq \emptyset)\},
$


For each MLLM block $m_j$ with non-empty $\mathcal{I}_j$, we refine its bounding box by taking the union of all qualifying UIED blocks: $B_j = \bigcup_{u_i \in \mathcal{I}_j} u_i,$ where the union operation computes the minimum bounding box that encompasses all blocks in $\mathcal{I}_j$. The final set of refined semantic-aware UI blocks is denoted as: $\mathcal{B} = \{B_1, B_2, \ldots, B_n\},$ where each $B_j$ represents a semantically meaningful UI component with precise boundaries.

\subsection{Visual-aware Graph-based Block Merge}
\label{subsec:VGBM}

\scheme\ merges UI blocks sharing similar visual appearances and UI structures for component-based implementation, enabling component code reuse within and across webpages. Given a set of webpages $\mathcal{P}={P_1,\ldots,P_n}$, where each page $P_k$ contains UI blocks $\mathcal{B}_k\subseteq\mathcal{B}$, our objective is to partition all blocks into groups $\mathcal{G}={G_1,\ldots,G_l}$ such that blocks within each group can be implemented as reusable components.




\subsubsection{Visual Similarity Metric}
To quantify the visual similarity between UI blocks, we compute a weighted combination of two metrics as $S_{\text{visual}}(B_i, B_j)$: $0.15 \cdot \text{SSIM}(B_i, B_j) + 0.85 \cdot (1 - \text{LPIPS}(B_i, B_j))$, where SSIM~\cite{wang2004image} (Structural Similarity Index) captures perceptual similarity, and LPIPS~\cite{zhang2018perceptual} (Learned Perceptual Image Patch Similarity) evaluates deep feature-based similarity. This metric serves as a prerequisite filter for the graph-based structural UI block grouping.


\subsubsection{Graph-based Block Grouping}
We employ a graph-based approach to identify UI blocks with similar structural layouts. This method captures both visually and structurally similar blocks, even though they have moderate visual differences.

We formally define a \textbf{UI Structure Graph (USG)} as an undirected attributed graph $G = (V, E)$ that captures the spatial relationships among UI elements within a block, where $V = \{v_1, v_2, \ldots, v_n\}$ is the set of nodes, with each node representing a detected UI element such as text, icon, button, or container, $E \subseteq V \times V \times \mathcal{T}$ is the set of undirected edges, with each edge $(v_i, v_j, t) \in E$ representing a spatial relationship of type $t \in \mathcal{T}$ between nodes $v_i$ and $v_j$. 
The edge type set $\mathcal{T} = \mathcal{T}_S \cup \mathcal{T}_A$ encompasses spatial relations $\mathcal{T}_S = \{\text{Vertical}, \text{Horizontal}, \text{NW-SE}, \text{NE-SW}\}$ and alignment relations $\mathcal{T}_A = \{\text{Align-Top}, \text{Align-Bottom}, \text{Align-Left}, \text{Align-Right}\}$. NW-SE denotes diagonal from northwest to southeast (top-left to bottom-right), and  NE-SW denotes diagonal from northeast to southwest (top-right to bottom-left).





\textbf{Graph Construction.} For each UI block $B_i$, we construct its corresponding graph $G(B_i)$ by first detecting all internal UI elements using UIED~\cite{xie2020uied}, then establishing edges based on their spatial configurations. We create two types of edges: (1) \textit{Spatial edges} determined by the relative positions of node center points, e.g., a vertical edge when nodes are vertically aligned based on their center coordinates; (2) \textit{Alignment edges} determined by whether element boundaries coincide, e.g., an ``Align-Left'' edge when the left boundaries of two elements align. Detailed edge creation criteria are provided in Appendix~\ref{append:graph_construct}

\textbf{Merging Criteria.} We merge UI blocks $B_i$ and $B_j$ when they satisfy both visual similarity and structural similarity conditions:
\begin{equation}
S_{\text{visual}}(B_i, B_j) > t_{\text{visual}} \land \text{GraphMatch}(G(B_i), G(B_j)),
\end{equation}
where $t_{\text{visual}}$ is the visual similarity threshold, and $\text{GraphMatch}\\(G_i, G_j)$ returns true if one of the following structural conditions holds: (1) $G_i$ and $G_j$ are isomorphic graphs, indicating identical structural layouts; (2) $G_i$ is a subgraph of $G_j$ (or vice versa), indicating one block contains the complete structure of another; (3) $G_i$ and $G_j$ share a common subgraph of significant size, indicating partial structural similarity. During graph matching, two edges can be matched only if they represent the same type of spatial and alignment relationship, while two nodes can be matched when their size similarity exceeds a predefined threshold $t_\text{size}$.

The visual similarity threshold $t_{\text{visual}} = 0.5$ ensures that blocks with moderate to high visual similarity are considered for structural matching, capturing blocks with similar structures but varying visual details (e.g., different text content or color schemes) while filtering out dissimilar blocks. The size similarity threshold $t_{\text{size}} = 0.7$ allows matching between nodes with comparable dimensions while tolerating reasonable size variations, accommodating blocks that maintain structural proportions despite absolute size differences. 

Through this visual-aware graph-based merging process, we obtain the final component groups $\mathcal{G} = \{G_1, G_2, \ldots, G_l\}$, where each group $G_i$ contains UI blocks sharing both visual appearance and structural consistency, enabling their implementation as reusable components across webpages.

\subsection{Component-based Code Generation}
\label{subsec:CCG}

After forming component groups $\mathcal{G} = {G_1, \dots, G_l}$, we employ a hierarchical code generation strategy. First, all UI block regions in the screenshot are masked, and the MLLM generates the overall webpage template with placeholders. For each group $G_i$, the MLLM generates reusable component code and instance-specific UI block code; unique blocks are generated individually. The generated code is integrated with the template via placeholders, producing a complete webpage with reusable components, improving modularity, and reducing computation. Prompt details are in Appendix~\ref{subsec:prompt_detail}.


\subsection{Priority-based Element-wise Feedback}
\label{subsec:PEF}

Existing feedback mechanisms are coarse-grained, feeding the generated and ground-truth UIs directly into an MLLM for self-refinement. Such feedback lacks explicit modification instructions (e.g., ``move the title to right'', ``expand the sign in button''), making it difficult for MLLMs to localize erroneous elements, while also introducing unnecessary token overhead by processing the entire UI. In contrast, benefiting from our component-based design, \scheme\ performs element-wise comparisons within UI blocks, enabling precise and lightweight refinements.




\begin{algorithm}[ht]
\footnotesize
\caption{Element Matching Algorithm}
\label{alg:element_matching}
\begin{algorithmic}[1]
\REQUIRE Ground truth UI elements $E_{gt} = \{e_1^{gt}, e_2^{gt}, ..., e_m^{gt}\}$
\REQUIRE Generated UI elements $E_{gen} = \{e_1^{gen}, e_2^{gen}, ..., e_n^{gen}\}$
\REQUIRE CLIP model for visual similarity
\ENSURE Matched elements $\mathcal{M}$
\STATE $\mathcal{M} \leftarrow \emptyset$ \COMMENT{Matched pairs}
\STATE $\mathcal{M}_{gen} \leftarrow \emptyset$ \COMMENT{Matched generated elements}

\FOR{each $e_i^{gt} \in E_{gt}$}
    \STATE $bestMatch \leftarrow NULL$, $maxSim \leftarrow 0$
    
    \FOR{each $e_j^{gen} \in E_{gen}$}
        \IF{$e_j^{gen} \notin \mathcal{M}_{gen}$}
            \IF{$hasText(e_i^{gt})$ \AND $hasText(e_j^{gen})$}
                \STATE $sim \leftarrow TextSimilarity(e_i^{gt}.text, e_j^{gen}.text)$
            \ELSE
                \STATE $sim_{pos} \leftarrow 1 - \frac{||(e_i^{gt}.x, e_i^{gt}.y) - (e_j^g.x, e_j^g.y)||}{diagonal}$
                \STATE $sim_{size} \leftarrow 1 - \frac{||e_i^{gt}.size - e_j^{gen}.size||}{max(e_i^{gt}.size, e_j^{gen}.size)}$
                \STATE $sim_{clip} \leftarrow CLIP(e_i^{gt}.image, e_j^{gen}.image)$
                \STATE $sim \leftarrow \alpha \cdot sim_{pos} + \beta \cdot sim_{size} + \gamma \cdot sim_{clip}$
            \ENDIF
            
            \IF{$sim > maxSim$}
                \STATE $maxSim \leftarrow sim$, $bestMatch \leftarrow e_j^g$
            \ENDIF
        \ENDIF
    \ENDFOR
    
    \IF{$bestMatch \neq NULL$ \AND $maxSim > \theta$}
        \STATE $\mathcal{M} \leftarrow \mathcal{M} \cup \{(e_i^{gt}, bestMatch)\}$
        \STATE $\mathcal{M}_{gen} \leftarrow \mathcal{M}_{gen} \cup \{bestMatch\}$
    \ENDIF
\ENDFOR



\RETURN $\mathcal{M}$
\end{algorithmic}
\end{algorithm}

\subsubsection{UI Blocks Ranking}
Not all UI blocks exhibit errors of equal severity. To optimize the refinement process, we prioritize blocks with more severe and visually salient issues. Specifically, we first compute the similarity between each generated block and its corresponding ground-truth block. To account for visual prominence, we weight each block’s priority by its area proportion relative to the entire UI: $\text{priority}_{\text{score}}(b_i) =
-\left(\text{sim}(b_i^{\text{gen}}, b_i^{\text{gt}})\right)
\times (1-\frac{\text{area}(b_i)}{\text{area}(\text{UI})}).$
Blocks with lower similarity (indicating more significant errors) and larger visual presence receive higher priority scores. We then rank all blocks and refine the top-$K$ UI blocks.

\subsubsection{Element Matching Strategy}

After selecting the UI blocks for refinement, we perform element matching within each block. We first apply UIED~\cite{xie2020uied} to detect UI elements and extract their properties such as position, size, and textual content. The detected elements from the generated and ground-truth UIs are then matched to establish correspondences. We design a matching algorithm that pairs each element in a ground-truth UI block with its counterpart in the corresponding generated UI block. Algorithm~\ref{alg:element_matching} summarizes the matching procedure. For text-containing elements, we compute text similarity using edit distance (Lines 8--10). For non-text elements (e.g., icons, images, and containers), we compute a composite similarity score that integrates spatial position, size, and CLIP-based visual features (Lines 11--15).

\subsubsection{Feedback Generation.}

After matching, we analyze each pair for inconsistencies in text, size, position, and visual through the similarity calculation shown in Lines 8-12. Unmatched ground truth elements trigger addition instructions, while unmatched generated elements trigger deletion instructions. For matched pairs with discrepancies, we generate specific modification instructions to provide the MLLM with actionable guidance for precise refinement. These instructions are categorized into four types:
\begin{itemize}[leftmargin=*]
    \item \textit{Element addition}: When ground truth elements have no matching counterparts in the generated UI (e.g., ``Add a login button at position (100, 50) with size 80×30'').
    \item \textit{Element deletion}: When generated elements do not correspond to any ground truth elements (e.g., ``Delete the redundant navigation bar'').
    \item \textit{Position adjustment}: When spatial coordinates differ significantly (e.g., ``Move the title 20px to the right'').
    \item \textit{Size modification}: When dimensions do not match (e.g., ``Expand the sign-in button width to 120px'').
    \item \textit{Content update}: When text content is incorrect or missing (e.g., ``Change button text from `Submit' to `Sign In''').
\end{itemize}
The detailed translation process is shown in Appendix~\ref{sec:feedback_instruction} and the detailed prompt is shown in Appendix~\ref{subsec:prompt_detail}. These fine-grained instructions provide the MLLM with actionable guidance for refinement, enabling more precise fixes than coarse-grained feedback.

\section{Experiment Setup}

\subsection{Metrics}

\textbf{High-level Visual Metrics.} Evaluate overall webpage appearance: (1) CLIP score~\cite{radford2021learning} measures visual similarity via CLIP-ViT-B/32 embeddings; (2) SSIM~\cite{wang2004image} assesses perceptual similarity in luminance, contrast, and structure, reflecting layout consistency. \textbf{Low-level Visual Metrics.} To capture fine-grained differences, we adopt element-matching metrics from Design2Code~\cite{si2024design2code}. Visual blocks are detected and aligned via the Jonker-Volgenant algorithm, then evaluated by: (1) Block-match: matched area ratio; (2) Text similarity: character overlap (Sørensen-Dice); (3) Color similarity: CIEDE2000 perceptual difference; (4) Position similarity: block center alignment accuracy.

\textbf{Code Metrics.}  (1) Tree BLEU~\cite{gui2025webcode2m}: it measures the structural similarity of the HTML DOM tree by calculating the recall of 1-height subtrees relative to the reference. (2) Repetitive Ratio: the percentage of lines of code containing duplicate elements out of the total lines of code. (3) Reuse Rate: the fraction of repeated regions that are captured by reusable components, computed against the reference.

\textbf{Component Related Metrics.} 
We evaluate component-level segmentation via one-to-one matching based on IoU. Each ground-truth component is matched to the predicted component with the highest IoU, yielding True Positives (TP), False Negatives (FN), and False Positives (FP), from which Precision, Recall, and F1 are computed. We also report the mean IoU over matched pairs as an overall quality measure. For component merging, we assess clustering consistency using Adjusted Rand Index (ARI)~\cite{santos2009use}, which corrects for chance agreement and is invariant to label permutations, and V-measure~\cite{rosenberg2007v}, which evaluates the harmonic mean of homogeneity and completeness. Details of metrics are shown in Appendix~\ref{subsec:metrics}.

\subsection{Baselines}

\textbf{Open-source Fine-tuned Models.} These methods fine-tune vision–language models on domain-specific datasets for image-to-code generation: (1) WebCoder-1.3B~\cite{gui2025webcode2m}, fine-tuned on WebCode2M using Pix2Struct-1.3B; (2) WebSight VLM-8B~\cite{laurençon2024unlocking}, trained on WebSight with DoRA; (3) Design2Code-18B~\cite{si2024design2code}, fine-tuned on WebSight using CogAgent with LoRA.

\textbf{Prompt-based Methods.} These approaches rely on carefully designed prompts to guide MLLMs in end-to-end code generation. (1) \textbf{Design2Code}~\cite{si2024design2code} proposes three prompt strategies: direct prompts for straightforward generation, text-augmented prompts that incorporate additional textual descriptions, and self-refinement prompts that enable iterative improvement of generated code. (2) \textbf{Component-based Prompt} (Appendix~\ref{subsec:prompt_detail}): We design a component-based prompt to guide the MLLM in generating both reusable components and the final webpage code in a single pass. 


\textbf{Segmentation-based Methods.} These methods first decompose complex webpages into manageable segments, then generate code for each segment before assembling the final output. (1) \textbf{UICopilot}~\cite{gui2025uicopilot} predicts the DOM tree structure for segmentation and synthesizes UI blocks through hierarchical generation. (2) \textbf{DCGen}~\cite{wan2025divide} detects dividing lines to partition webpages into segments, generates code for each segment independently, and merges them into the complete webpage. (3) \textbf{LaTCoder}~\cite{gui2025latcoder} identifies grid-aligned dividing lines to create structured blocks, employs Layout-as-Thought prompting for block-wise generation, and assembles results via absolute positioning or MLLM-based fusion. (4) \textbf{LayoutCoder}~\cite{wu2025mllm} constructs element relationships and parses layout trees to capture complex UI structures, then generates layout-preserved code through guided fusion techniques.


For prompt-based and segmentation-based methods, we employ Gemini-2.5-Pro~\cite{comanici2025gemini} as the backbone MLLM. Following common practice, we set the temperature to 0 for deterministic generation and the maximum output token limit to 16,384 to accommodate complex webpage structures. We use vue framework for component-based implementation.

\subsection{Research Questions}


\begin{itemize}[leftmargin=*]

    \item (RQ1) \textbf{Performance and Reusability}. Compared with other methods, does \scheme \ achieve better performance on complex webpages while generating reusable code?


    
    \item (RQ2) \textbf{Ablation study}. How do different parts contribute to \scheme's performance?
    
    \item (RQ3) \textbf{Parameter study}. How do key parameters affect the performance of \scheme?
    
    \item (RQ4) \textbf{Case Study.} Why does \scheme \ work?
\end{itemize}

\section{Experiment Results}

\begin{table*}[h]
\centering
\caption{The overall performance of \scheme \ and baseline models. Because only \scheme and component-based prompting produce component-based code, the reuse rate for all other methods is zero. Bold values indicate the optimal performance, and underlined values indicate the second best performance.}
\label{tab:overall_performance}
\resizebox{\linewidth}{!}{
\renewcommand{\arraystretch}{0.5}
\begin{tabular}{@{}l|cccc|cc|ccc@{}}
\toprule
\textbf{Method} 
& \multicolumn{4}{c|}{\textbf{Low-level Metrics}} 
& \multicolumn{2}{c|}{\textbf{High-level Metrics}} 
& \multicolumn{3}{c}{\textbf{Code Metrics}} \\ 
\cmidrule{2-10}
& Block-Match & Text & Position & Color 
& CLIP & SSIM 
& TreeBLEU & Repetitive(\%) & Reuse(\%) \\ 
\midrule
\multicolumn{10}{c}{\textit{Open-source Fine-tuned Models}} \\ 
\midrule
WebCoder-1.3B (WWW'24) 
& 0.1171 & 0.0919 & 0.0667 & 0.0772 
& 0.5244 & 0.3688 
& 0.1400 & 20.97\% & --- \\
WebSight-8B (Arixv'24) 
& 0.1346 & 0.1003 & 0.0837 & 0.0955 
& 0.6551 & 0.5255 
& 0.0322 & 31.62\% & --- \\
Design2Code-18B (NAACL'25) 
& 0.1763 & 0.1450 & 0.1049 & 0.1061 
& 0.7056 & 0.5169 
& 0.0609 & 19.51\% & --- \\
\midrule
\multicolumn{10}{c}{\textit{Prompt-based Methods}} \\ 
\midrule
Design2Code (Direct Prompt) 
& 0.8524 & 0.8403 & 0.7417 & 0.5531 
& 0.8134 & 0.6812 
& 0.3407 & 22.78\% & --- \\
Design2Code (Text-Augmented) 
& 0.8494 & 0.8416 & 0.7465 & 0.5847 
& 0.8231 & 0.6834 
& 0.3294 & 25.15\% & --- \\
Design2Code (Self-Revision) 
& 0.7509 & 0.7439 & 0.6509 & 0.6486 
& 0.8194 & 0.6427 
& 0.3421 & 24.35\% & --- \\
Component-based Prompt 
& 0.7899 & 0.7779 & 0.6834 & 0.6930 
& 0.8018 & 0.6536 
& 0.3619 & \textbf{0.46\%} & \underline{55.23\%} \\
\midrule
\multicolumn{10}{c}{\textit{Segmentation-based Methods}} \\ 
\midrule
UICopilot (WWW'24) 
& 0.6501 & 0.6428 & 0.5687 & 0.5856 
& 0.7799 & 0.6000 
& 0.2986 & 27.92\% & --- \\
DCGen (FSE'25) 
& 0.8295 & 0.8219 & 0.7269 & 0.7394 
& \underline{0.8244} & \textbf{0.7128} 
& 0.3855 & 48.50\% & --- \\
LatCoder (KDD'25) 
& 0.7588 & 0.7535 & 0.6791 & 0.6854 
& 0.6735 & 0.5735 
& 0.3276 & 39.85\% & --- \\
LayoutCoder (ISSTA'25) 
& \underline{0.8742} & \underline{0.8641} & \underline{0.7776} & \underline{0.7959} 
& 0.7801 & 0.6311 
& \textbf{0.3974} & 57.43\% & --- \\
\midrule
\textbf{\scheme\ (Ours)} 
& \textbf{0.9301} & \textbf{0.9221} & \textbf{0.8172} & \textbf{0.8334} 
& \textbf{0.8327} & \underline{0.7021} 
& \underline{0.3958} & \underline{2.10\%} & \textbf{59.96\%} \\
\bottomrule
\end{tabular}}
\end{table*}

\subsection{Performance and Reusablity (RQ1)}

\textbf{Automatic Metrics.} As shown in Table~\ref{tab:overall_performance}, \scheme\ achieves the best performance across all evaluation dimensions. For low-level visual metrics, \scheme\ shows substantial improvements in Block-Match, text, position, and color accuracy, thanks to Hybrid Semantic-aware Block Segmentation (HSBS) for precise UI block detection and Priority-based Element-wise Feedback (PEF) for enhancing fine-grained visual consistency. For high-level visual metrics, it attains the highest CLIP and SSIM scores, reflecting effective UI template code generation and layout fidelity. Regarding code metrics, \scheme\ achieves the highest Tree BLEU.

\textbf{Human Evaluation.} To complement automatic metrics, we conduct human evaluation on one-third of the dataset following Design2Code~\cite{si2024design2code} (details in Appendix~\ref{subsec:human_eval}). Figure~\ref{fig:human_eval} shows pairwise preference results, where five annotators compare each method against direct prompting using majority voting. Open-source fine-tuned models are excluded due to substantially lower performance. Higher win rates and lower loss rates indicate superior code quality. The results confirm that \scheme\ consistently outperforms baselines, validating the reliability of our automatic metrics.

\begin{figure}[ht]
    \centering
    \includegraphics[width = .35\textwidth]{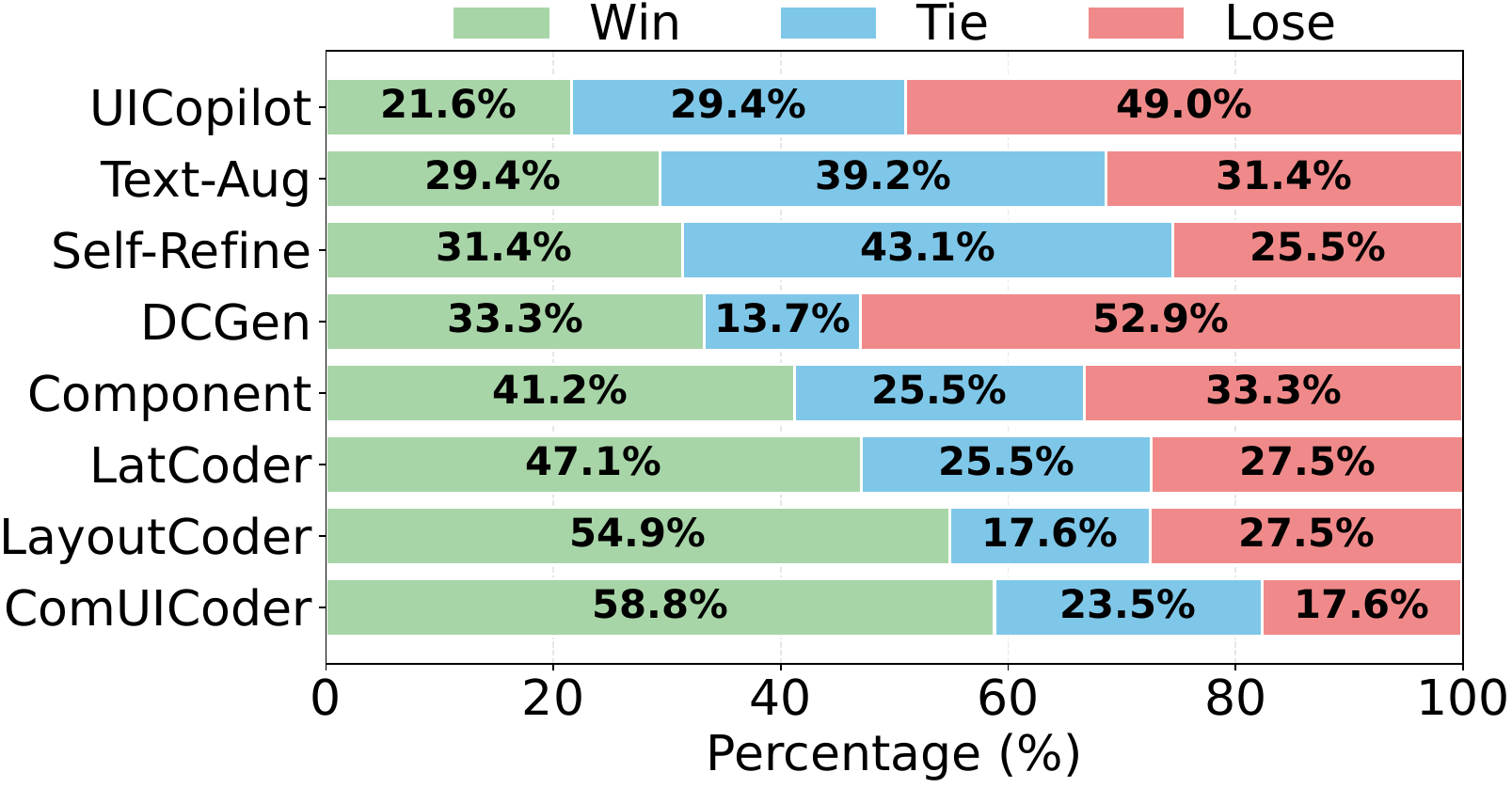}
    \caption{Human Evaluation Results. Component denotes component-based prompting.}
    \label{fig:human_eval}
\end{figure}

\textbf{Code Reusability.} Table~\ref{tab:overall_performance} shows that \scheme\ achieves the lowest repetitive ratio, indicating its tendency to explicitly define reusable components rather than repeatedly generating structurally similar code fragments. This demonstrates that the proposed Visual-aware Graph-based Block Merge (VGBM) effectively promotes component reuse, resulting in more modular and maintainable code. We further compare reuse rates across methods, where \scheme \ attains the higher reuse rate compare with direct component-based prompting, demonstrating the effectiveness of its component-based design and the proposed VGBM merge strategy.






\subsection{Ablation Study (RQ2)}

\subsubsection{Ablation of the three main modules.}

\begin{table}[h]
\caption{The performance of 5 variants for \scheme.}
\label{tab:ablation}
\resizebox{\linewidth}{!}{
\setlength{\tabcolsep}{.15em}{
\renewcommand{\arraystretch}{0.5}  
\begin{tabular}{@{}cccc|cccccc@{}}
\toprule
HSBS & VGBM & PEF & & Block-Match & CLIP & TreeBLEU & Reuse Rate \\
\midrule
{\color[HTML]{FF0000} X} & {\color[HTML]{FF0000} X} & {\color[HTML]{FF0000} X} & $C_0$ & 0.7478 & 0.8186 & 0.4682 & 57.96\%\\
{\color[HTML]{FF0000} X} & {\color[HTML]{FF0000} X} & {\color[HTML]{00CD66} Y} & $C_1$ & 0.7917 & 0.8261 & 0.4571 & 57.96\%\\
{\color[HTML]{00CD66} Y} & {\color[HTML]{FF0000} X} & {\color[HTML]{00CD66} Y} & $C_2$ & 0.8605 & 0.8478 & 0.4939 & 0\% \\
{\color[HTML]{00CD66} Y} & {\color[HTML]{00CD66} Y} & {\color[HTML]{FF0000} X} & $C_3$ & 0.8845 & 0.8416 & 0.4987 & \textbf{67.25\%}  \\
{\color[HTML]{00CD66} Y} & {\color[HTML]{00CD66} Y} & {\color[HTML]{00CD66} Y} & $C_4$ & \textbf{0.8978} & \textbf{0.8453} & \textbf{0.5039} & \textbf{67.25\%} \\
\bottomrule
\end{tabular}}}
\end{table}

\begin{figure*}[h]
    \centering
    \includegraphics[width = .95\textwidth]{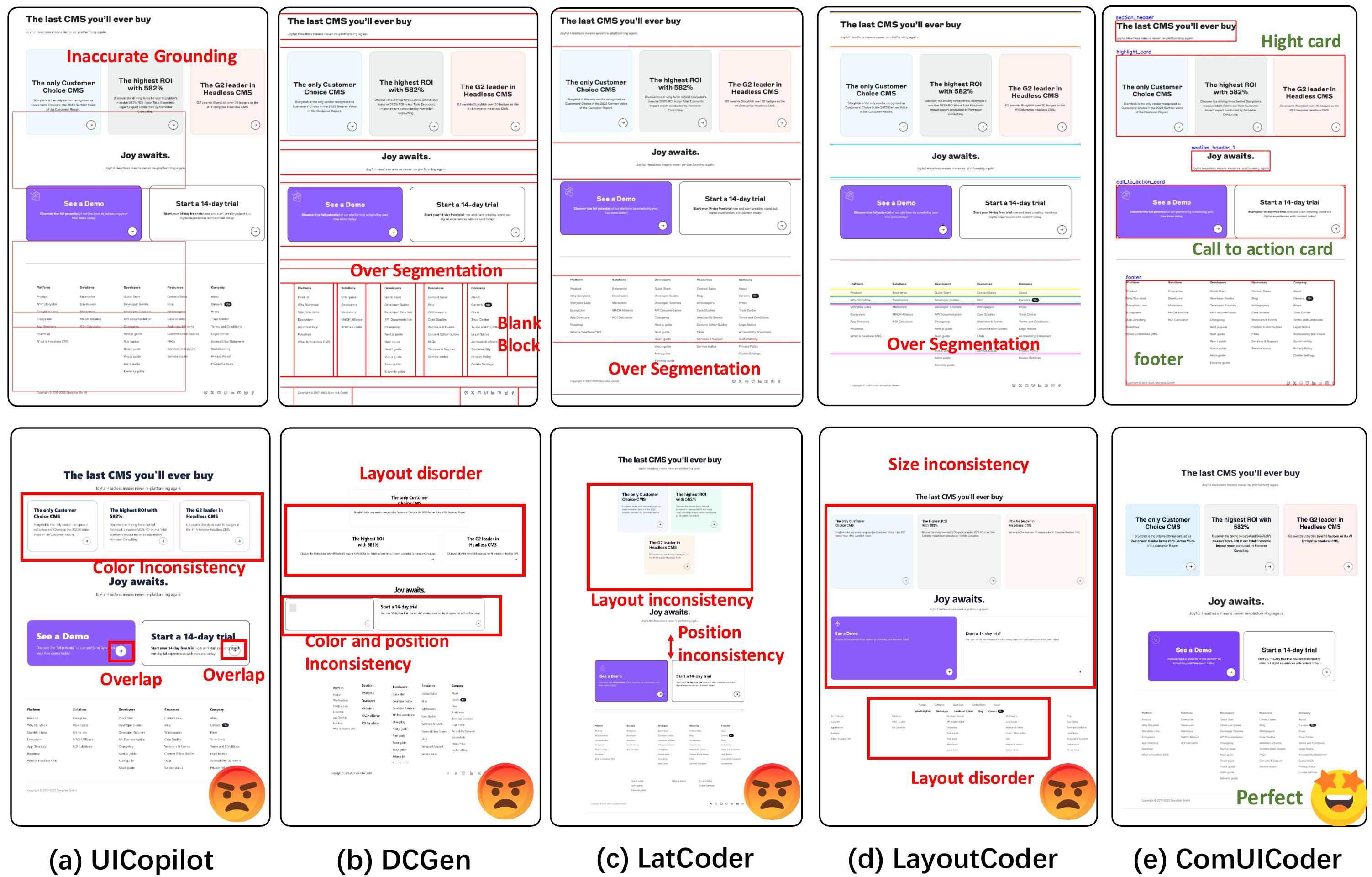}
    \caption{Case study. The first row is the segmentation results and the second row is the generation results.}
    \label{fig:case_study}
\end{figure*}

\scheme\ comprises three core components: Hybrid Semantic-aware Block Segmentation (\textbf{HSBS}), Visual-aware Graph-based Block Merge (\textbf{VGBM}), and Priority-based Element-wise Feedback (\textbf{PEF}). We conduct an ablation study with five variants of \scheme\ ($C_0$--$C_5$) to evaluate the contribution of each component, where {\color[HTML]{00CD66}Y} and {\color[HTML]{FF0000}X} denote inclusion and removal, respectively. Since VGBM depends on HSBS, removing HSBS also disables VGBM. Configuration $C_5$ corresponds to the complete \scheme. As shown in Table~\ref{tab:ablation}, each component contributes positively to overall performance. The reuse rates of $C_0$ and $C_1$, as well as $C_3$ and $C_4$, are identical because feedback does not affect component planning, while $C_2$ yields zero reuse due to independent block implementation after removing VGBM. The full configuration achieves the best performance, demonstrating the complementary effects of all components.

\subsubsection{The Segmentation Method Deep Dive.} We conduct an in-depth comparison of different segmentation methods. Table~\ref{tab:ablation_seg} reveals three key findings: (1) \scheme  \ demonstrates the best component detection performance with the highest IoU and F1 score significantly outperforming all baseline methods. (2) \scheme \ generates a moderate number of blocks (2,298) with zero blank blocks, effectively avoiding the over-segmentation problem that affects other methods. (3) Neither UIED alone nor pure MLLM detection achieves optimal performance, while their combination (\scheme)  yields the best results, which validates the rationality of our hybrid design.



\begin{table}[ht]
\caption{Semantic UI block detection performance compare.}
\resizebox{\linewidth}{!}{
\setlength{\tabcolsep}{.15em}{
\renewcommand{\arraystretch}{0.5}
\begin{tabular}{@{}l|cccccc@{}}
\toprule
Methods & Block & Blank Block  & Acc & Pre & F1 & IoU \\ \midrule
UICopilot       & \textbf{1125} & \underline{5} & 0.4521 & \textbf{0.8951} & 0.6227 & 0.1427 \\
DCGen       & 3691 & 44 & 0.2945 & 0.3002 & 0.4550 & 0.4428 \\
LatCoder    & 2626 & 86 & 0.4818 & 0.5796 & 0.6503 & 0.3318 \\
LayoutCoder  & 5615 & 169 & 0.2911 & 0.2967 & 0.4509 & 0.3685 \\

UIED  & 33861 & 280 & 0.0607   & 0.0607   & 0.1144   & 0.2969   \\
MLLM        & 2578 & 15 & \underline{0.6792}   & 0.7269   & \underline{0.8090}   & \underline{0.5653}   \\
\scheme       & \underline{2298} & \textbf{0} & \textbf{0.7075}   & \underline{0.7836}   & \textbf{0.8287}   & \textbf{0.6076}   \\ \bottomrule
\end{tabular}}}
\label{tab:ablation_seg}
\end{table}

\subsubsection{The Merging Method Deep Dive.}
We evaluate different strategies for merging UI blocks into cohesive groups. Table~\ref{tab:ablation_merge} shows that neither Visual-based Merging nor Graph-based Merging alone achieves optimal performance, while our VGBM method, which combines both visual similarity and structural relationships, achieves the best overall performance.

\begin{table}[ht]
\caption{UI blocks merging performance comparison.}
\resizebox{\linewidth}{!}{
\renewcommand{\arraystretch}{0.5}
\setlength{\tabcolsep}{.15em}{
\begin{tabular}{@{}l|cccc@{}}
\toprule
Methods & Homogeneity & Completeness  & V-Measure & ARI \\ \midrule
Graph-based Merging  & \textbf{0.9963} & 0.7573 & 0.8573 & 0.1935 \\
Visual-basd Merging  & 0.9471 & 0.9251 & 0.9349   & 0.7202   \\
VGBM (Ours)        & 0.9611 & \textbf{0.9261} & \textbf{0.9424}   & \textbf{0.7567}   \\\bottomrule
\end{tabular}}}
\label{tab:ablation_merge}
\end{table}

\begin{figure}[h]
    \subfigure[Detection Metrics.]{
    \centering
    \includegraphics[width = .2\textwidth]{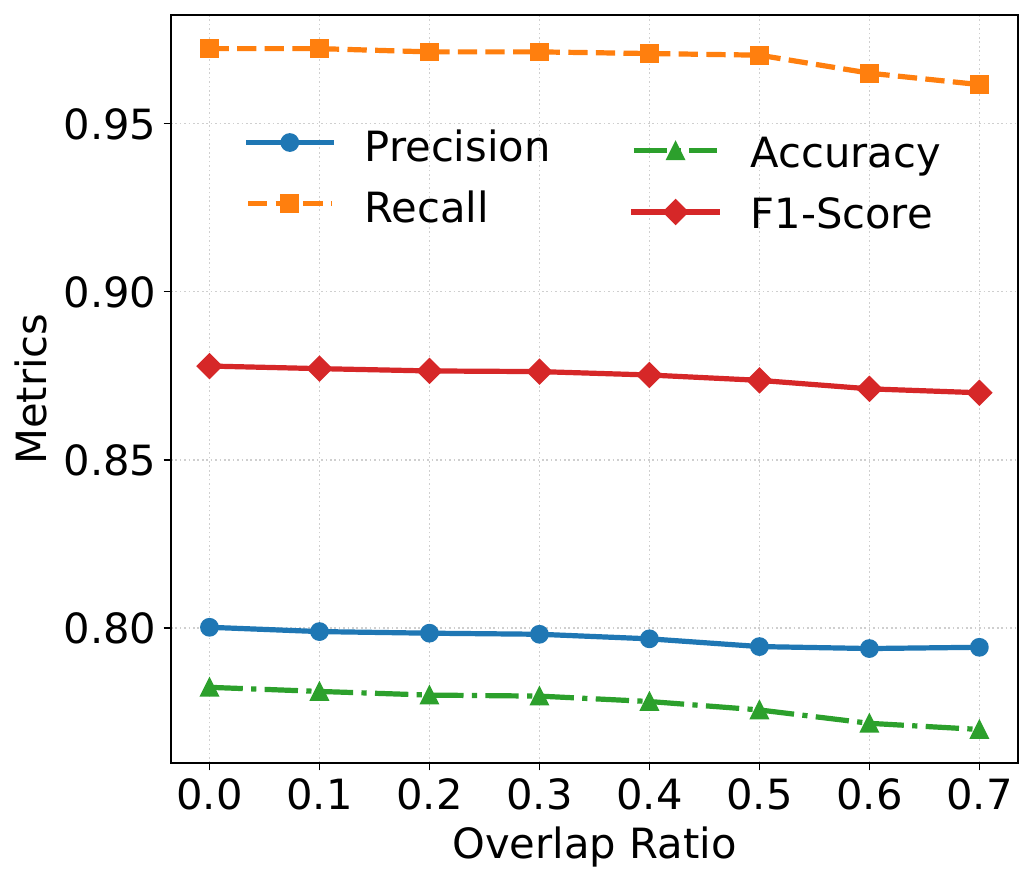}
    }
    \subfigure[IoU Score.]{
    \centering
    \includegraphics[width = .2\textwidth]{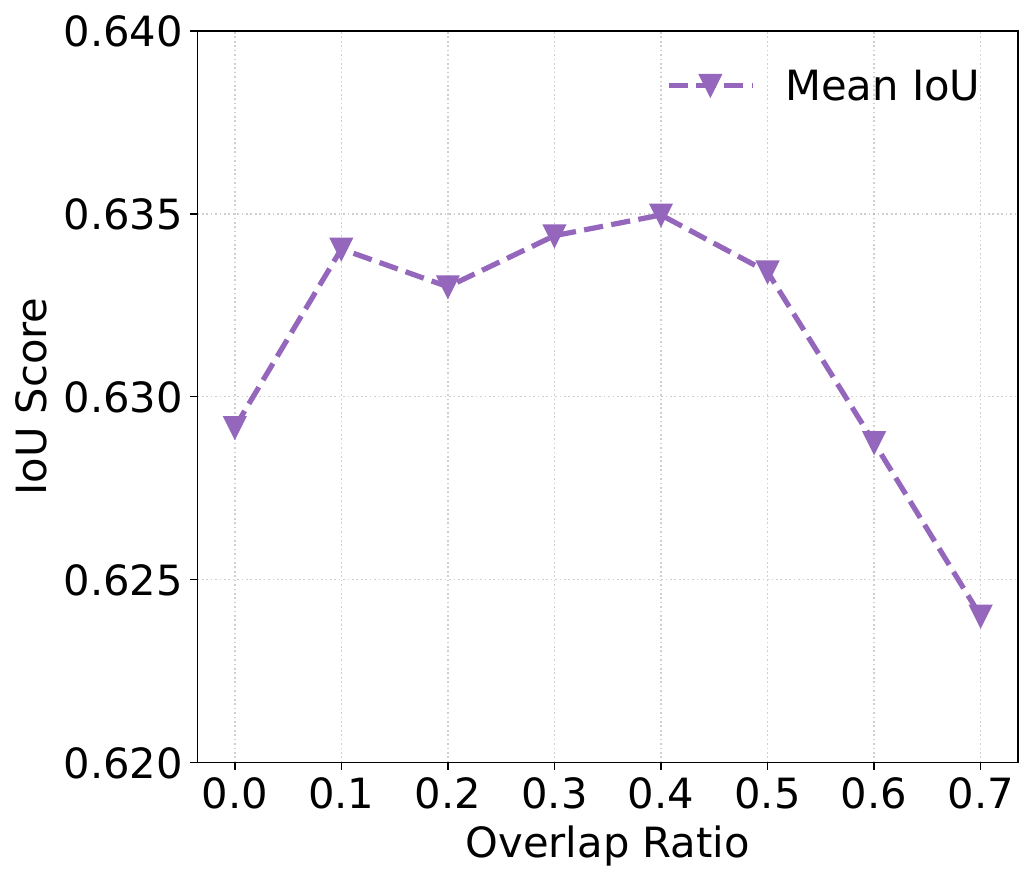}
    }
    \caption{Parameters of HSBS.}
    \label{fig:para_seg}
\end{figure}

\begin{figure}[h]
    \subfigure[Adjusted Rand Index.]{
    \centering
    \includegraphics[width = .2\textwidth]{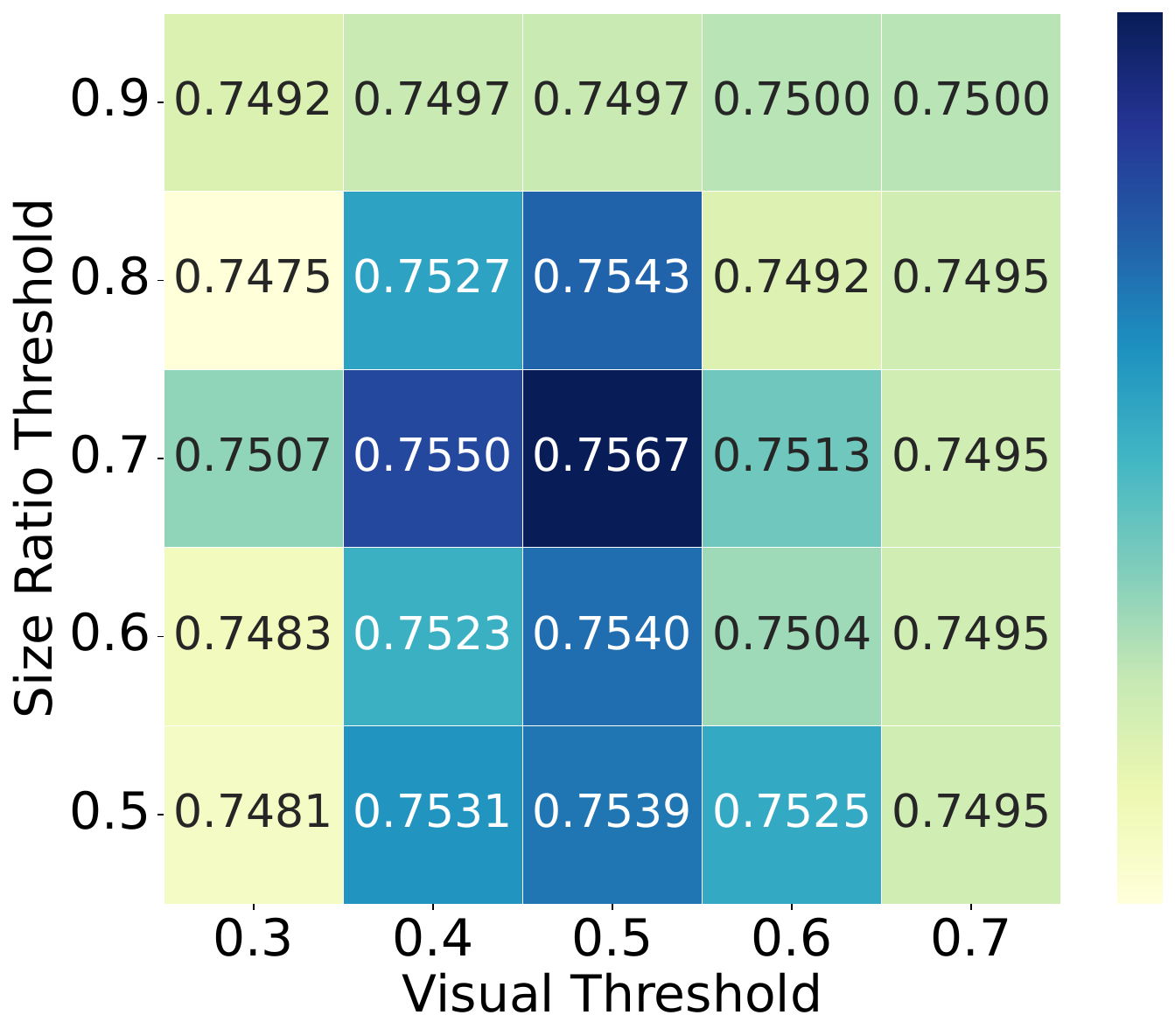}
    }
    \subfigure[V-Measure.]{
    \centering
    \includegraphics[width = .2\textwidth]{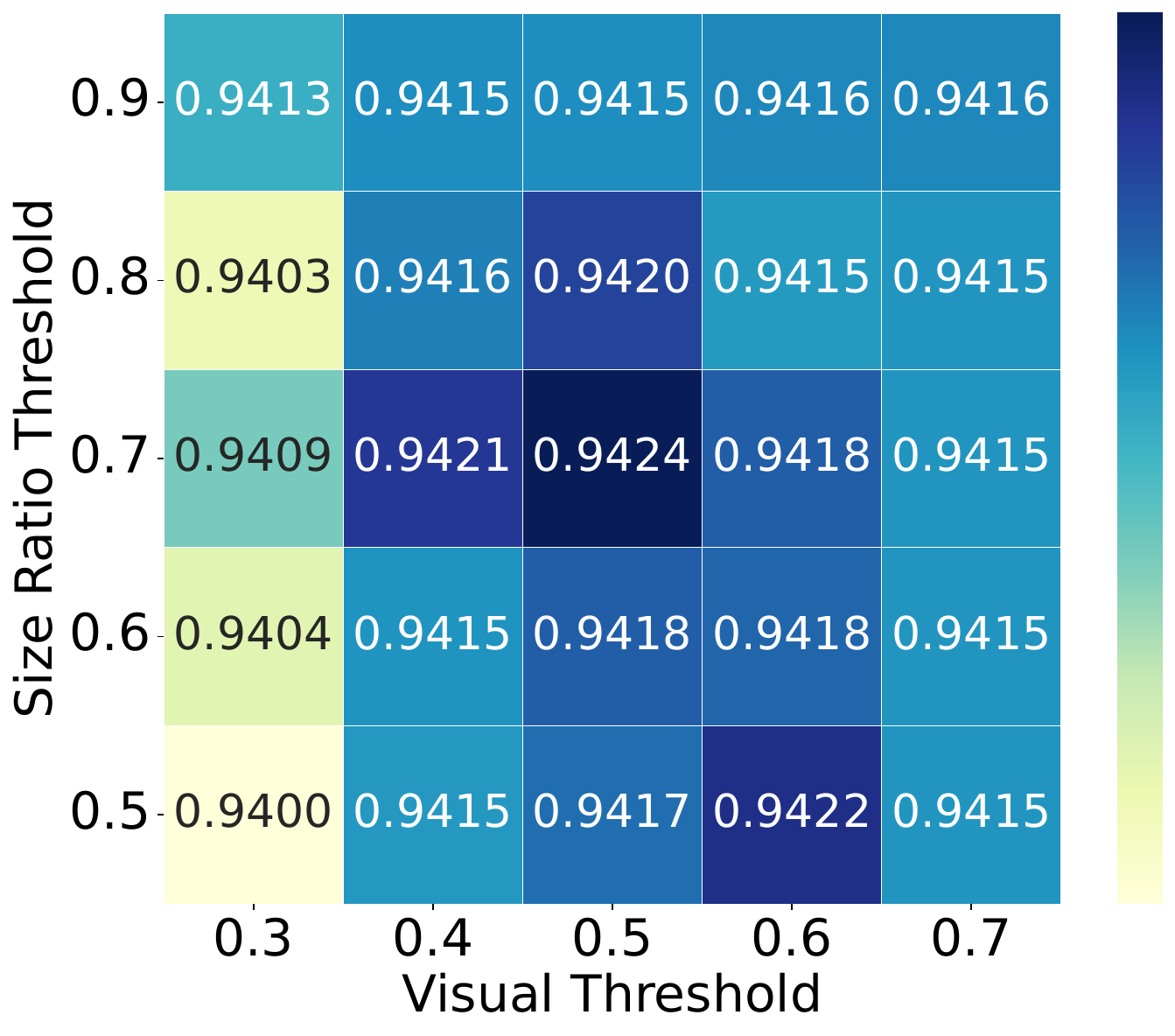}
    }
    \caption{Parameters of Visual-aware Graph-based Merging.}
    \label{fig:merge_para}
\end{figure}

\subsection{Parameter Analysis (RQ3)}



\subsubsection{The overlap threshold $t_{overlap}$ of segmentation.} Figure~\ref{fig:para_seg} illustrates the impact of the overlap ratio threshold on HSBS performance. All detection metrics remain relatively stable across different overlap ratios, demonstrating robustness. IoU score peaks at $t_{overlap}=0.4$, suggesting that a moderate overlap threshold optimally balances UIED's precise boundaries with the MLLM's semantic regions.


\subsubsection{The threshold $t_{visual}$ and $t_{size}$ of the merging.} Figure~\ref{fig:merge_para} shows that VGBM achieves optimal performance when the visual threshold is set to 0.5 and the structure threshold is around 0.7, indicating that moderate visual similarity combined with high structural similarity ensures precise merging of similart UI blocks.











\subsection{Case Study (RQ4)}

We sample a webpage from \benchmark\ for a case study; due to its complexity, only a portion is shown. The first row of Figure.~\ref{fig:case_study} compares segmentation results across methods. UICopilot yields inaccurate bounding boxes, while DCGen, LatCoder, and LayoutCoder tend to over-fragment semantically coherent UI blocks (e.g., the footer) or produce meaningless regions (e.g., blank blocks). In contrast, \scheme\ accurately identifies semantically coherent blocks.
The second row of Figure.~\ref{fig:case_study} shows qualitative generation results. Baseline methods exhibit noticeable inconsistencies in layout, size, color, and element positioning, whereas \scheme\ largely alleviates these issues by leveraging high-quality HSBS segmentation and fine-grained PEF feedback. Further code inspection shows that only \scheme\ explicitly defines reusable components (e.g., feature cards, call-to-action cards, and the footer) and reuses them across the page, demonstrating its superior ability to promote code reuse. Additional cases are provided in Appendix~\ref{appendix:case_study}.

\section{Conclusion}

We investigate key limitations of MLLMs in complex UI-to-code generation, including fragmented segmentation, redundant code, and cross-page inconsistencies. We introduce \benchmark, a multi-page benchmark with component annotations, and propose \scheme, a component-based framework integrating semantic-aware segmentation, component merging, and element-level refinement. Extensive experiments show that \scheme \ significantly improves code reusability, consistency, and overall generation quality on complex websites, highlighting the importance of component-centric reasoning for scalable UI code generation.

\balance

\bibliography{sample-base}

@article{wan2025divide,
  title={Divide-and-Conquer: Generating UI Code from Screenshots},
  author={Wan, Yuxuan and Wang, Chaozheng and Dong, Yi and Wang, Wenxuan and Li, Shuqing and Huo, Yintong and Lyu, Michael},
  journal={Proceedings of the ACM on Software Engineering (FSE 2025)},
  volume={2},
  number={FSE},
  pages={2099--2122},
  year={2025},
  publisher={ACM New York, NY, USA}
}

@article{yun2024web2code,
  title={Web2code: A large-scale webpage-to-code dataset and evaluation framework for multimodal llms},
  author={Yun, Sukmin and Thushara, Rusiru and Bhat, Mohammad and Wang, Yongxin and Deng, Mingkai and Wang, Jinhong and Tao, Tianhua and Li, Junbo and Li, Haonan and Nakov, Preslav and others},
  journal={Advances in neural information processing systems (NeurIPS 2024)},
  volume={37},
  pages={112134--112157},
  year={2024}
}

@article{wu2025mllm,
  title={MLLM-Based UI2Code Automation Guided by UI Layout Information},
  author={Wu, Fan and Gao, Cuiyun and Li, Shuqing and Wen, Xin-Cheng and Liao, Qing},
  journal={Proceedings of the ACM on Software Engineering (ISSTA 2025)},
  volume={2},
  pages={1123--1145},
  year={2025},
  publisher={ACM New York, NY, USA}
}

@inproceedings{gui2025webcode2m,
  title={Webcode2m: A real-world dataset for code generation from webpage designs},
  author={Gui, Yi and Li, Zhen and Wan, Yao and Shi, Yemin and Zhang, Hongyu and Chen, Bohua and Su, Yi and Chen, Dongping and Wu, Siyuan and Zhou, Xing and others},
  booktitle={Proceedings of the ACM on Web Conference (WWW 2025)},
  pages={1834--1845},
  year={2025}
}

@inproceedings{gui2025latcoder,
  title={Latcoder: Converting webpage design to code with layout-as-thought},
  author={Gui, Yi and Li, Zhen and Zhang, Zhongyi and Wang, Guohao and Lv, Tianpeng and Jiang, Gaoyang and Liu, Yi and Chen, Dongping and Wan, Yao and Zhang, Hongyu and others},
  booktitle={Proceedings of the 31st ACM SIGKDD Conference on Knowledge Discovery and Data Mining (KDD 2025)},
  pages={721--732},
  year={2025}
}

@article{xiao2024interaction2code,
      title={Interaction2Code: Benchmarking MLLM-based Interactive Webpage Code Generation from Interactive Prototyping}, 
      author={Jingyu Xiao and Yuxuan Wan and Yintong Huo and Zixin Wang and Xinyi Xu and Wenxuan Wang and Zhiyao Xu and Yuhang Wang and Michael R. Lyu},
      journal={arXiv preprint arXiv:2411.03292},
      year={2025}
}

@article{wan2024mrweb,
  title={Mrweb: An exploration of generating multi-page resource-aware web code from ui designs},
  author={Wan, Yuxuan and Dong, Yi and Xiao, Jingyu and Huo, Yintong and Wang, Wenxuan and Lyu, Michael R},
  journal={arXiv preprint arXiv:2412.15310},
  year={2024}
}

@article{xiao2025designbench,
  title={Designbench: A comprehensive benchmark for mllm-based front-end code generation},
  author={Xiao, Jingyu and Wang, Ming and Lam, Man Ho and Wan, Yuxuan and Liu, Junliang and Huo, Yintong and Lyu, Michael R},
  journal={arXiv preprint arXiv:2506.06251},
  year={2025}
}

@article{xiao2025efficientuicoder,
  title={Efficientuicoder: Efficient mllm-based ui code generation via input and output token compression},
  author={Xiao, Jingyu and Zhang, Zhongyi and Wan, Yuxuan and Huo, Yintong and Liu, Yang and Lyu, Michael R},
  journal={arXiv preprint arXiv:2509.12159},
  year={2025}
}

@article{dang2025envisioning,
  title={Envisioning Future Interactive Web Development: Editing Webpage with Natural Language},
  author={Dang, Truong Hai and Xiao, Jingyu and Huo, Yintong},
  journal={arXiv preprint arXiv:2510.26516},
  year={2025}
}

@article{liu2025benchmarking,
  title={Benchmarking MLLM-based Web Understanding: Reasoning, Robustness and Safety},
  author={Liu, Junliang and Xiao, Jingyu and Tang, Wenxin and Wang, Wenxuan and Wang, Zhixian and Zhang, Minrui and Yu, Shuanghe},
  journal={arXiv preprint arXiv:2509.21782},
  year={2025}
}

@inproceedings{tang2025slidecoder,
    title = "{S}lide{C}oder: Layout-aware {RAG}-enhanced Hierarchical Slide Generation from Design",
    author = "Tang, Wenxin  and
      Xiao, Jingyu  and
      Jiang, Wenxuan  and
      Xiao, Xi  and
      Wang, Yuhang  and
      Tang, Xuxin  and
      Li, Qing  and
      Ma, Yuehe  and
      Liu, Junliang  and
      Tang, Shisong  and
      Lyu, Michael R.",
    editor = "Christodoulopoulos, Christos  and
      Chakraborty, Tanmoy  and
      Rose, Carolyn  and
      Peng, Violet",
    booktitle = "Proceedings of the 2025 Conference on Empirical Methods in Natural Language Processing (EMNLP 2025)",
    month = nov,
    year = "2025",
    address = "Suzhou, China",
    publisher = "Association for Computational Linguistics",
    url = "https://aclanthology.org/2025.emnlp-main.458/",
    doi = "10.18653/v1/2025.emnlp-main.458",
    pages = "9026--9050",
    ISBN = "979-8-89176-332-6"
}

@article{gao2025treat,
  title={TREAT: A Code LLMs Trustworthiness/Reliability Evaluation and Testing Framework},
  author={Gao, Shuzheng and Li, Eric John and Lam, Man Ho and Xiao, Jingyu and Wan, Yuxuan and Wang, Chaozheng and Tik, Ng Man and Lyu, Michael R},
  journal={arXiv preprint arXiv:2510.17163},
  year={2025}
}

@inproceedings{santos2009use,
  title={On the use of the adjusted rand index as a metric for evaluating supervised classification},
  author={Santos, Jorge M and Embrechts, Mark},
  booktitle={International conference on artificial neural networks},
  pages={175--184},
  year={2009},
  organization={Springer}
}

@misc{moz_top500,
  title        = {Moz Top 500},
  author       = {{Moz}},
  year         = {2026},
  howpublished = {\url{https://moz.com/top500}},
  note         = {Accessed: 2026-2}
}

@inproceedings{rosenberg2007v,
  title={V-measure: A conditional entropy-based external cluster evaluation measure},
  author={Rosenberg, Andrew and Hirschberg, Julia},
  booktitle={Proceedings of the 2007 joint conference on empirical methods in natural language processing and computational natural language learning (EMNLP-CoNLL)},
  pages={410--420},
  year={2007}
}

@article{comanici2025gemini,
  title={Gemini 2.5: Pushing the frontier with advanced reasoning, multimodality, long context, and next generation agentic capabilities},
  author={Comanici, Gheorghe and Bieber, Eric and Schaekermann, Mike and Pasupat, Ice and Sachdeva, Noveen and Dhillon, Inderjit and Blistein, Marcel and Ram, Ori and Zhang, Dan and Rosen, Evan and others},
  journal={arXiv preprint arXiv:2507.06261},
  year={2025}
}

@inproceedings{radford2021learning,
  title={Learning transferable visual models from natural language supervision},
  author={Radford, Alec and Kim, Jong Wook and Hallacy, Chris and Ramesh, Aditya and Goh, Gabriel and Agarwal, Sandhini and Sastry, Girish and Askell, Amanda and Mishkin, Pamela and Clark, Jack and others},
  booktitle={International conference on machine learning},
  pages={8748--8763},
  year={2021},
  organization={PMLR}
}

@inproceedings{gui2025uicopilot,
  title={UICoPilot: Automating UI synthesis via hierarchical code generation from webpage designs},
  author={Gui, Yi and Wan, Yao and Li, Zhen and Zhang, Zhongyi and Chen, Dongping and Zhang, Hongyu and Su, Yi and Chen, Bohua and Zhou, Xing and Jiang, Wenbin and others},
  booktitle={Proceedings of the ACM on Web Conference (WWW 2025)},
  pages={1846--1855},
  year={2025}
}

@article{wang2004image,
  title={Image quality assessment: from error visibility to structural similarity},
  author={Wang, Zhou and Bovik, Alan C and Sheikh, Hamid R and Simoncelli, Eero P},
  journal={IEEE transactions on image processing},
  volume={13},
  number={4},
  pages={600--612},
  year={2004},
  publisher={IEEE}
}

@article{wan2025automatically,
  title={Automatically Generating Web Applications from Requirements Via Multi-Agent Test-Driven Development},
  author={Wan, Yuxuan and Liang, Tingshuo and Xu, Jiakai and Xiao, Jingyu and Huo, Yintong and Lyu, Michael R},
  journal={arXiv preprint arXiv:2509.25297},
  year={2025}
}

@inproceedings{xie2020uied,
  title={UIED: a hybrid tool for GUI element detection},
  author={Xie, Mulong and Feng, Sidong and Xing, Zhenchang and Chen, Jieshan and Chen, Chunyang},
  booktitle={Proceedings of the 28th ACM Joint Meeting on European Software Engineering Conference and Symposium on the Foundations of Software Engineering (ASE 2020)},
  pages={1655--1659},
  year={2020}
}

@misc{uimockup,
    title = {UI Mockups},
    url = {https://www.uxpin.com/studio/blog/what-is-a-mockup-the-final-layer-of-ui-design/},
    author = {UI-Mockup},
    month = {January},
    year = {2025}
}

@inproceedings{acsirouglu2019automatic,
  title={Automatic HTML code generation from mock-up images using machine learning techniques},
  author={A{\c{s}}{\i}ro{\u{g}}lu, Batuhan and Mete, B{\"u}{\c{s}}ta R{\"u}meysa and Y{\i}ld{\i}z, Eyy{\"u}p and Nal{\c{c}}akan, Ya{\u{g}}{\i}z and Sezen, Alper and Da{\u{g}}tekin, Mustafa and Ensari, Tolga},
  booktitle={2019 Scientific Meeting on Electrical-Electronics \& Biomedical Engineering and Computer Science (EBBT)},
  pages={1--4},
  year={2019},
  organization={Ieee}
}

@article{chen2022code,
  title={Code generation from a graphical user interface via attention-based encoder--decoder model},
  author={Chen, Wen-Yin and Podstreleny, Pavol and Cheng, Wen-Huang and Chen, Yung-Yao and Hua, Kai-Lung},
  journal={Multimedia Systems},
  volume={28},
  number={1},
  pages={121--130},
  year={2022},
  publisher={Springer}
}

@article{moran2018machine,
  title={Machine learning-based prototyping of graphical user interfaces for mobile apps},
  author={Moran, Kevin and Bernal-C{\'a}rdenas, Carlos and Curcio, Michael and Bonett, Richard and Poshyvanyk, Denys},
  journal={IEEE Transactions on Software Engineering},
  volume={46},
  number={2},
  pages={196--221},
  year={2018},
  publisher={IEEE}
}

@article{cizotto2023web,
  title={Web pages from mockup design based on convolutional neural network and class activation mapping},
  author={Cizotto, Andr{\'e} Armstrong Janino and de Souza, Rodrigo Clemente Thom and Mariani, Viviana Cocco and dos Santos Coelho, Leandro},
  journal={Multimedia Tools and Applications},
  volume={82},
  number={25},
  pages={38771--38797},
  year={2023},
  publisher={Springer}
}

@article{jain2019sketch2code,
  title={Sketch2Code: transformation of sketches to UI in real-time using deep neural network},
  author={Jain, Vanita and Agrawal, Piyush and Banga, Subham and Kapoor, Rishabh and Gulyani, Shashwat},
  journal={arXiv preprint arXiv:1910.08930},
  year={2019}
}

@inproceedings{nguyen2015reverse,
  title={Reverse engineering mobile application user interfaces with remaui (t)},
  author={Nguyen, Tuan Anh and Csallner, Christoph},
  booktitle={2015 30th IEEE/ACM International Conference on Automated Software Engineering (ASE)},
  pages={248--259},
  year={2015},
  organization={IEEE}
}

@inproceedings{beltramelli2018pix2code,
  title={pix2code: Generating code from a graphical user interface screenshot},
  author={Beltramelli, Tony},
  booktitle={Proceedings of the ACM SIGCHI symposium on engineering interactive computing systems},
  pages={1--6},
  year={2018}
}

@inproceedings{zhang2018perceptual,
  title={The Unreasonable Effectiveness of Deep Features as a Perceptual Metric},
  author={Zhang, Richard and Isola, Phillip and Efros, Alexei A and Shechtman, Eli and Wang, Oliver},
  booktitle={CVPR},
  year={2018}
}

@inproceedings{si2024design2code,
  title={Design2Code: Benchmarking multimodal code generation for automated front-end engineering},
  author={Si, Chenglei and Zhang, Yanzhe and Li, Ryan and Yang, Zhengyuan and Liu, Ruibo and Yang, Diyi},
  booktitle={Proceedings of the 2025 Conference of the Nations of the Americas Chapter of the Association for Computational Linguistics: Human Language Technologies (NAACL 2025)},
  volume={1},
  pages={3956--3974}
}

@misc{laurençon2024unlocking,
      title={Unlocking the conversion of Web Screenshots into HTML Code with the WebSight Dataset}, 
      author={Hugo Laurençon and Léo Tronchon and Victor Sanh},
      year={2024},
      eprint={2403.09029},
      archivePrefix={arXiv},
      primaryClass={cs.HC}
}

@article{Xu2021Image2e,
  author = {Y. Xu and L. Bo and X. Sun and B. Li and J. Jiang and W. Zhou},
  title = {image2emmet: Automatic code generation from web user interface image},
  journal = {Journal of Software: Evolution and Process},
  year = {2021},
  volume = {33},
  number = {8},
  pages = {e2369}
}

@inproceedings{Chen2018FromUI,
  author = {C. Chen and T. Su and G. Meng and Z. Xing and Y. Liu},
  title = {From UI design image to GUI skeleton: a neural machine translator to bootstrap mobile GUI implementation},
  booktitle = {Proceedings of the 40th International Conference on Software Engineering},
  year = {2018},
  pages = {665--676}
}
\bibliographystyle{ACM-Reference-Format}

\newpage


\appendix
\section{Appendices}

\subsection{Details of Data Collection and Annotation}
\label{subsec:data_collection_detail}

We introduce \benchmark, a novel UI-to-code benchmark that advances beyond existing single-page datasets by comprising multi-subpage samples with fine-grained component-level annotations for comprehensive evaluation of complex and reusable web code generation. The benchmark construction employs systematic collection and annotation methodologies to capture diverse real-world design patterns.

\textbf{Annotator Recruitment.} We employ five PhD students majoring in computer science, each with more than two years of professional experience in front-end development.

\textbf{Website Selection.} Annotators are instructed to select websites from the Moz Top 500 global rankings~\cite{moz_top500}, which represent the most visited and influential websites worldwide. To ensure dataset quality and diversity, we establish comprehensive selection criteria that prioritize: (1) \textit{design complexity}: websites must exhibit rich, multi-component layouts with at least 10 distinct component types; (2) \textit{component reusability}: design patterns should recur across multiple pages within the site; (3) \textit{structural diversity}: websites must contain multiple page templates (e.g., homepage, listing pages, detail pages); and (4) \textit{domain variety}: the selection spans e-commerce, news media, social platforms, technology services, and entertainment sites to capture diverse design paradigms. For each website, annotators identify 2-5 representative subpages that collectively showcase the site's component ecosystem while avoiding redundant page templates. Detailed selection guidelines are provided in our repo \url{https://github.com/WebPAI/ComUICoder}.

\textbf{Semantic UI Blocks Annotation.} A semantic UI block is a self-contained, visually distinct UI region on a webpage that serves a specific function. It is characterized by four key properties: (1) \textit{clear boundaries}, with identifiable visual separation from surrounding elements; (2) \textit{functional purpose}, supporting user interaction or information display; (3) \textit{structural integrity}, existing as a complete unit rather than a fragment; and (4) \textit{reusability potential}, meaning it can be reused elsewhere on the site. Annotators follow a structured annotation protocol with three key principles:  (1) \textit{boundary completeness}: each annotation encompasses all visual elements, padding, and margins that constitute the functional unit. (2) \textit{grouping consecutive repetitions}: when identical components appear consecutively (e.g., product cards in a grid), they are annotated as a single group to facilitate loop-based code generation; (3) \textit{outermost-only annotation}: for nested elements, only the outermost container is annotated to maintain component integrity and simplify implementation;

\textbf{UI Blocks Grouping.} UI blocks are grouped across pages based on structural, visual, and and functional similarity to identify reusable design patterns within each website. Grouping is governed by three criteria: (1) \textit{structural equivalence}, requiring identical element composition and layout hierarchy; (2) \textit{visual consistency}, including typography, color, spacing, and visual treatments; and (3) \textit{functional equivalence}, ensuring the same interaction purpose. The methodology supports mixed annotation granularities: group-level annotations with multiple instances and single-instance annotations are merged when their component structures match, ensuring accurate representation of design pattern reuse.

\begin{figure}[ht]
    \centering
    \includegraphics[width = .48\textwidth]{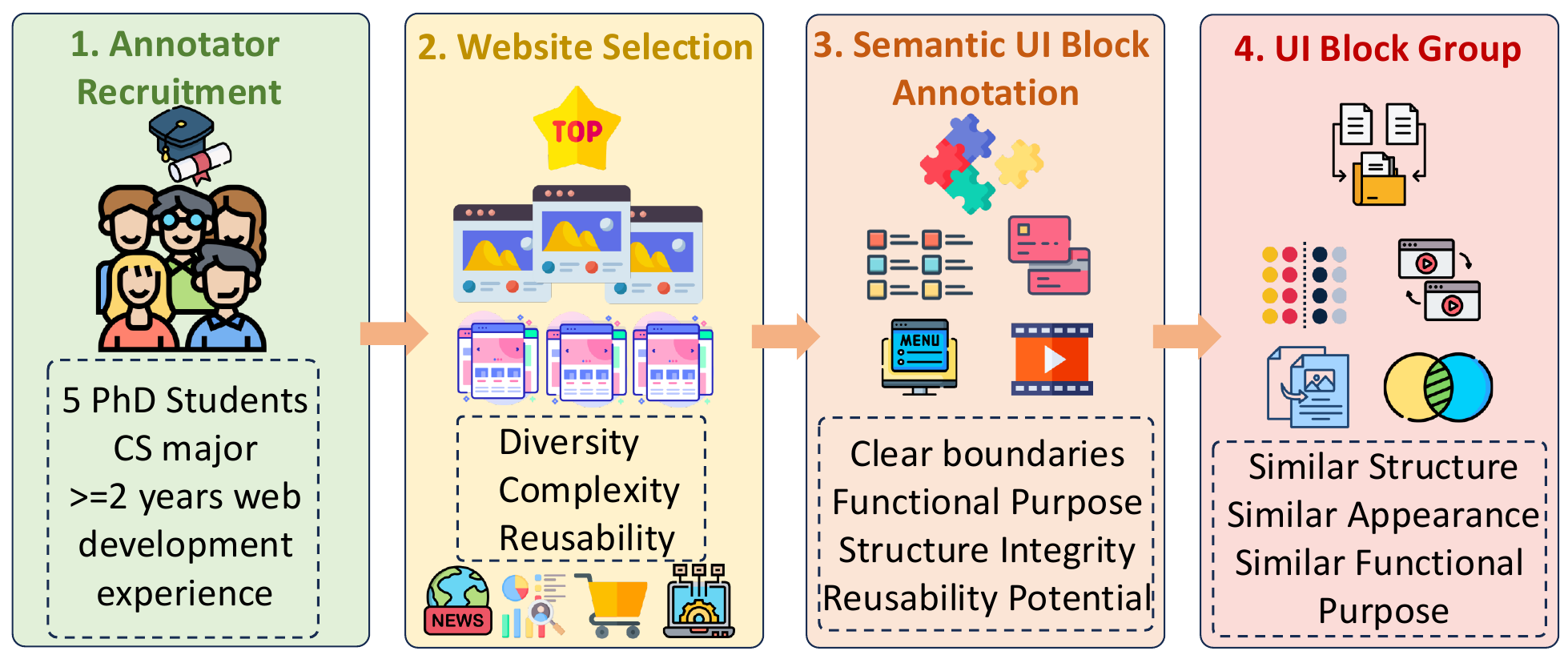}
    \caption{The process of \benchmark \  construction.}
    \label{fig:bench_construct}
\end{figure}

\subsection{Data statistics of \benchmark}
\label{appendix:data_stat}



Figure~\ref{fig:category_pie} shows that our benchmark covers a diverse range of web topics with more than 13 types, including technology, news, software and so on.

\begin{figure}[ht]
    \centering
    \includegraphics[width = .49\textwidth]{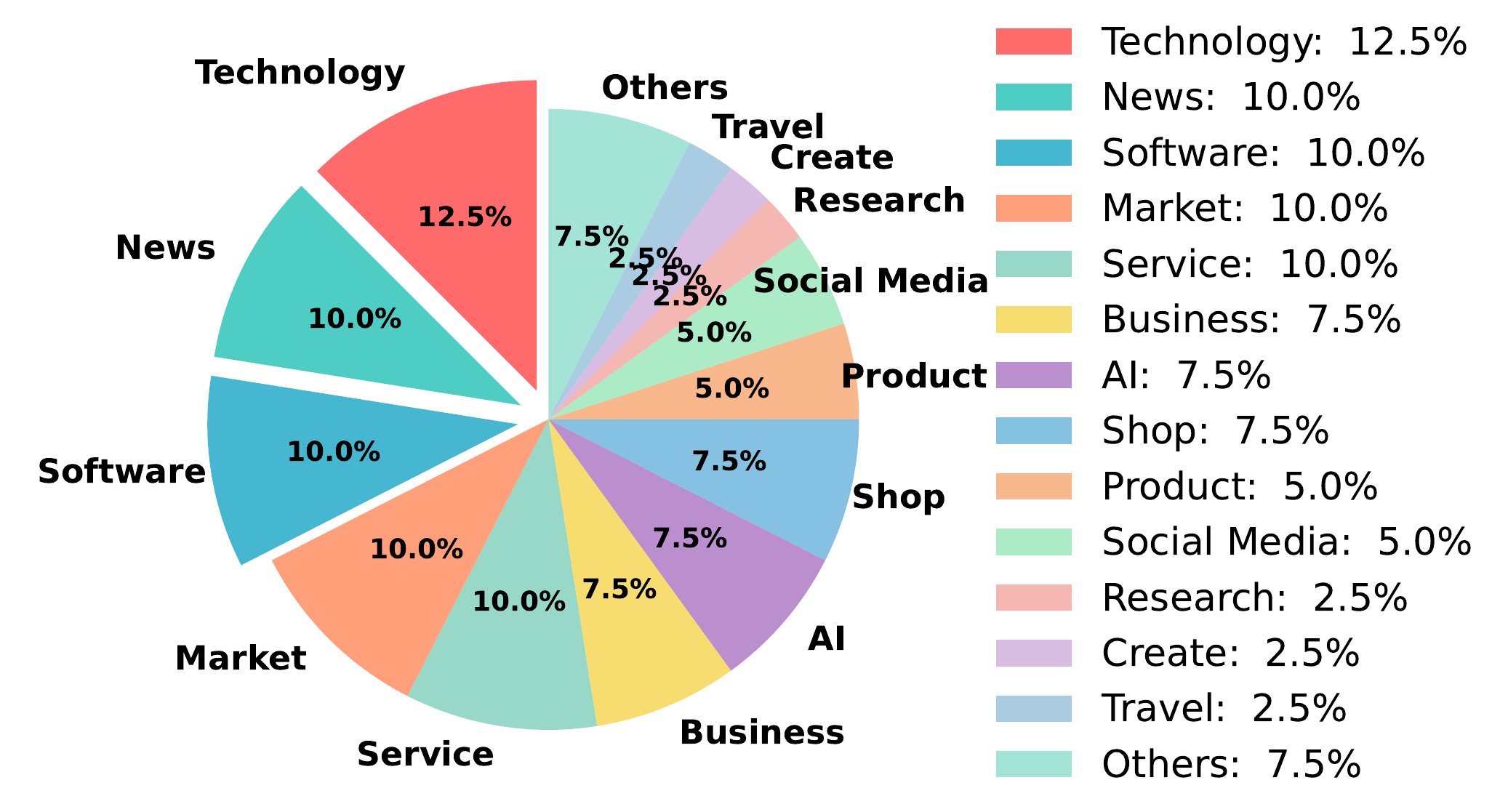}
    \caption{Website topic distribution of \benchmark.}
    \label{fig:category_pie}
\end{figure}



We define the following metrics to characterize component reusability from UI element quantity perspectives:

\begin{itemize}
\item \textbf{Subpage (A)}: The number of distinct subpages within a website sample, indicating the breadth of website coverage and potential for cross-page component reuse.

\item \textbf{Groups with $\geq$2 Components (B)}: The number of component groups containing at least two reusable components, measuring the presence of recurring design patterns that justify component abstraction.

\item \textbf{Total Groups (C)}: The total number of component groups identified across all subpages, representing the overall diversity of UI block types.

\item \textbf{Components in Groups with $\geq$2 Members (D)}: The total count of components belonging to groups with at least two members, quantifying actual reusable component instances that benefit from component-based implementation.

\item \textbf{Total Components (E)}: The total number of individual component instances across all websites, capturing the complete inventory of UI blocks.

\item \textbf{Reusability Ratio (F = B/C)}: The proportion of component groups that contain reusable patterns, indicating the efficiency of component grouping.

\item \textbf{Component Reuse Coverage (G = D/E)}: The proportion of total components belonging to reusable groups, measuring the overall reusability coverage of the website.
\end{itemize}

\textbf{Analysis}: Table~\ref{tab:web_level_stats} reveals significant component reusability potential in real-world websites. Our dataset shows a mean of 3.75 subpages per website with 9.93 reusable groups (out of 28.35 total groups), indicating that approximately 38\% of component groups exhibit genuine reuse patterns (mean F = 0.379). More importantly, the Component Reuse Coverage achieves a mean of 0.633, demonstrating that 63\% of all UI components are reusable—with some websites reaching near-complete componentization (max G = 0.9375). The mean of 32.95 reusable components per website (out of 51.38 total) provides ample instances for evaluating component generation quality. The substantial standard deviations across metrics (e.g., Std(G) = 0.1689) highlight diverse design practices in real-world websites, making our benchmark representative of varied scenarios. These statistics validate the practical importance of component-based code generation approaches, as the majority of UI elements in modern web design can benefit from reusable component implementations rather than hard-coded markup.

\begin{table}[h]
\centering
\caption{Reusability features statistics of \benchmark.}
\label{tab:web_level_stats}
\resizebox{0.95\linewidth}{!}{
\setlength{\tabcolsep}{.12em}{
\begin{tabular}{l|ccccccc}
\toprule
\multicolumn{8}{c}{Website-level}      \\ \midrule
 & A & B & C & D & E & F & G \\ \midrule
Mean & 3.75 & 9.93 & 28.35 & 32.95 & 51.38 & 0.3790 & 0.633 \\
Std & 0.62 & 3.63 & 9.11 & 13.94 & 15.73 & 0.1655 & 0.1689 \\
Min & 2 & 1 & 13 & 3 & 23 & 0.0345 & 0.0968 \\
Median & 4 & 10 & 28 & 33 & 50 & 0.3670 & 0.6614 \\
Max & 5 & 20 & 51 & 76 & 102 & 0.8333 & 0.9375 \\ \midrule
\multicolumn{8}{c}{Page-level}      \\ \midrule
Mean   &  -   & 7.21 & 12.13 & 8.79 & 13.7 & 0.6004 & 0.6345 \\
Std    &  -   & 3.03 & 3.47  & 4.03 & 4.1  & 0.2045 & 0.2087 \\
Min    &  -   & 1    & 4     & 1    & 4    & 0.0833 & 0.0833 \\
Median &  -   & 7    & 12    & 9    & 13   & 0.625  & 0.6667 \\
Max    &  -   & 15   & 23    & 20   & 28   & 1      & 1 \\ \bottomrule
\end{tabular}
}}
\end{table}

\subsection{Details of UI Structure Graph Construction}
\label{append:graph_construct}

\textbf{Graph Construction.} For a given UI block $B_i$, we construct its corresponding UI Structure Graph $G(B_i)$ by first detecting all internal UI elements using UIED~\cite{xie2020uied}, then establishing edges based on their spatial configurations.


\textit{Spatial Relationship Edges.} The spatial relationship type between two nodes $v_i$ and $v_j$ is determined by analyzing their relative positions, a spatial edge $(v_i, v_j, t_s) \in E$ with type $t_s \in \mathcal{T}_S$ is created when:
\begin{equation}
t_s = 
\begin{cases}
\text{Vertical} & \text{if } \Delta_y > \Delta_x \land \Delta_x < \tau \\
\text{Horizontal} & \text{if } \Delta_x > \Delta_y \land \Delta_y < \tau \\
\text{NW-SE} & \text{if } \theta \in [0°, 90°) \\
\text{NE-SW} & \text{if } \theta \in [90°, 180°)
\end{cases},
\end{equation}
where $\Delta_x = |\text{center}_x(v_j) - \text{center}_x(v_i)|$ and $\Delta_y = |\text{center}_y(v_j) - \text{center}_y(v_i)|$ denote the horizontal and vertical distances between node centers, $\tau$ is the alignment tolerance threshold, $\theta = \arctan(\Delta_y / \Delta_x)$ represents the angle between the connecting vector and the positive x-axis.



\textit{Alignment Relationship Edges.} Alignment edges capture precise layout constraints between elements that share common boundaries. An alignment edge $(v_i, v_j, t_a) \in E$ with type $t_a \in \mathcal{T}_A$ is created when:
\begin{equation}
t_a = 
\begin{cases}
\text{Align-Left} & \text{if } |x_{i1} - x_{j1}| < \epsilon \\
\text{Align-Right} & \text{if } |x_{i2} - x_{j2}| < \epsilon \\
\text{Align-Top} & \text{if } |y_{i1} - y_{j1}| < \epsilon \\
\text{Align-Bottom} & \text{if } |y_{i2} - y_{j2}| < \epsilon
\end{cases},
\end{equation}
where $(x_{i1}, y_{i1})$ and $(x_{i2}, y_{i2})$ represent the top-left and bottom-right corners of node $v_i$'s bounding box, and $\epsilon$ is the alignment precision tolerance.

\subsection{Translating the Inconsistency Issues to Natural Instructions}
\label{sec:feedback_instruction}




Our feedback generation pipeline consists of three stages: inconsistency detection, priority classification, and instruction synthesis.

\subsubsection{Inconsistency Detection and Classification}

For each matched element pair $(e^{gt}, e^{gen})$ identified in Algorithm~\ref{alg:element_matching}, we compute four types of similarity scores to detect potential mismatches:

\begin{itemize}[leftmargin=*]
    \item \textbf{Text Mismatch}: $sim_{text} = \text{EditDistance}(e^{gt}.text, e^{gen}.text)$ for text-containing elements
    \item \textbf{Size Mismatch}: 
    \begin{equation}
    \begin{aligned}
    diff_w &= \frac{|w^{gt} - w^{gen}|}{\max(w^{gt}, w^{gen})}, \quad diff_h = \frac{|h^{gt} - h^{gen}|}{\max(h^{gt}, h^{gen})} \\
    sim_{size} &= 1 - \frac{diff_w + diff_h}{2}
    \end{aligned}
    \end{equation}
    \item \textbf{Position Mismatch}: Using normalized coordinates,
    \begin{equation}
    sim_{pos} = 1 - \sqrt{(\bar{x}^{gt} - \bar{x}^{gen})^2 + (\bar{y}^{gt} - \bar{y}^{gen})^2}
    \end{equation}
    where $\bar{x} = x/W_{UI}$ and $\bar{y} = y/H_{UI}$ are normalized coordinates
    \item \textbf{Visual Mismatch}: $sim_{visual} = \text{CLIP}(e^{gt}.image, e^{gen}.image)$ for non-text elements
\end{itemize}
Each similarity score is compared against predefined thresholds to determine the severity level and mismatch type. We define three priority levels based on the deviation magnitude and visual importance:

\begin{equation}
priority\_level = 
\begin{cases}
\text{High} & \text{if } (1 - sim) > \tau_{high} \text{ and } area(e) > \tau_{area} \\
\text{Medium} & \text{if } \tau_{medium} < (1 - sim) \leq \tau_{high} \\
\text{Low} & \text{if } (1 - sim) \leq \tau_{medium}
\end{cases}
\end{equation}

where $\tau_{high}$, $\tau_{medium}$ are severity thresholds (e.g., 0.3, 0.15), and $\tau_{area}$ is the minimum area ratio for high-priority classification.

\subsubsection{Visual Reference Annotation}

To provide clear modification targets, the ground truth UI with reference markers (\texttt{ref\#1}, \texttt{ref\#2}, ...) and the generated UI with fix markers (\texttt{fix\#1}, \texttt{fix\#2}, ...) are annotated by the bounding box detected by UIED~\cite{xie2020uied}. Each marker corresponds to a matched element pair, creating a visual correspondence between the target state and current state.

\subsubsection{Natural Instruction Synthesis}

We translate detected issues into natural language using template-based generation:

\textbf{(1) Modify Instruction:} For matched pairs with detected issues:

\begin{itemize}
    \item \textbf{Position instructions}: 
    \begin{itemize}
        \item Move horizontally: ``move it \{left|right\}'' when $|x^{gt} - x^{gen}| > \epsilon$ and $|y^{gt} - y^{gen}| \leq \epsilon$
        \item Move vertically: ``move it \{up|down\}'' when $|y^{gt} - y^{gen}| > \epsilon$ and $|x^{gt} - x^{gen}| \leq \epsilon$
        \item Move diagonally: ``move it \{up|down\} and to the \{left|right\}'' when both dimensions differ
    \end{itemize}
    \item \textbf{Size instructions}:
    \begin{itemize}
        \item Resize width: ``make it \{narrower|wider\} (width: $w^{gen}$px → $w^{gt}$px)''
        \item Resize height: ``make it \{shorter|taller\} (height: $h^{gen}$px → $h^{gt}$px)''
        \item Resize both: ``make it \{larger|smaller\} (resize to $w^{gt} \times h^{gt}$px)''
    \end{itemize}
    \item \textbf{Text instructions}: ``change text to `$e^{gt}.text$''' (currently showing ``$e^{gen}.text$'')
    
    \item \textbf{Visual instructions}: ``adjust visual appearance to match reference'' for significant CLIP similarity drops
\end{itemize}
Multiple issues for the same element are concatenated with conjunctions (e.g., ``move it down and to the right, and make it narrower'').

\textbf{(2) Add Instruction}: For unmatched ground truth elements:
\begin{itemize}
    \item ``add \{element\_type\} at position ($x^{gt}$, $y^{gt}$) with size $w^{gt} \times h^{gt}$px''
    \item If text exists: ``add \{element\_type\} showing `$e^{gt}.text$' at position ($x^{gt}$, $y^{gt}$)''
    \item For visual elements: ``add \{element\_type\} matching ref\#i appearance''
\end{itemize}

\textbf{(3) Delete Instruction}: For unmatched generated elements:
\begin{itemize}
    \item ``remove the \{element\_type\} at position ($x^{gen}$, $y^{gen}$)''
    \item If text exists: ``remove the \{element\_type\} showing `$e^{gen}.text$'''
    \item For identifiable elements: ``delete the redundant element''
\end{itemize}

\subsubsection{Output Format}

We organize instructions into a structured format:

\begin{tcolorbox}[
    breakable,
    colback=gray!4,
    colframe=black!45,
    title={\centering\textbf{Structured Feedback Instructions}},
    fonttitle=\bfseries,
    boxrule=0.6pt,
    arc=2mm,
    left=1.2mm,
    right=1.2mm,
    top=1.0mm,
    bottom=1.0mm,
    before skip=2pt,
    after skip=2pt,
    skin=enhanced jigsaw,
    width=0.48\textwidth
]
\begin{verbatim}
[fix#1 → ref#1]: move it down, and make it taller 
(height: 40px → 51px) (currently showing "...")
[fix#2 → ref#2]: move it up and to the right 
(currently showing "...")
[fix#3 → ref#3]: make it slightly wider 
(width: 48px → 50px)

[add#1]: add elemment showing "Sign In" at position 
(120, 80) with size 100×40px
[add#2]: add icon at position (50, 150) matching 
ref#2 appearance

[del#1]: remove the redundant item showing "Extra Menu"
[del#2]: remove the redundant item at (200, 300)
\end{verbatim}

\end{tcolorbox}

This structured format, combined with the annotated UI images, provides the MLLM with clear, actionable guidance for iterative refinement. Each instruction references specific widgets through the fix/ref/add/del numbering system, enabling precise localization of modifications.

\subsection{Details of Metrics}
\label{subsec:metrics}

\subsubsection{Visual Metrics}

\paragraph{High-level Visual Metrics.}
High-level visual metrics assess the overall appearance fidelity between the generated webpage and the reference.

\textbf{CLIP Score.}
We use CLIP~\cite{radford2021learning} to measure semantic visual similarity. Given a reference image $I_r$ and a generated image $I_g$, we extract visual embeddings using CLIP-ViT-B/32 and compute the cosine similarity:
\begin{equation}
\text{CLIP}(I_r, I_g) = 
\frac{\langle \phi(I_r), \phi(I_g) \rangle}
{\|\phi(I_r)\| \, \|\phi(I_g)\|},
\end{equation}
where $\phi(\cdot)$ denotes the CLIP image encoder.

\textbf{SSIM.}
Structural Similarity Index Measure (SSIM)~\cite{wang2004image} evaluates perceptual similarity by comparing luminance, contrast, and structural information:
\begin{equation}
\text{SSIM}(I_r, I_g) =
\frac{(2\mu_r\mu_g + C_1)(2\sigma_{rg} + C_2)}
{(\mu_r^2 + \mu_g^2 + C_1)(\sigma_r^2 + \sigma_g^2 + C_2)},
\end{equation}
where $\mu$, $\sigma^2$, and $\sigma_{rg}$ denote mean, variance, and covariance.

\paragraph{Low-level Visual Metrics.}
High-level metrics alone cannot capture fine-grained discrepancies. We therefore adopt element-matching metrics from Design2Code~\cite{si2024design2code}.

Given reference and generated webpage screenshots, visual element blocks are detected and aligned using the Jonker--Volgenant assignment algorithm. Based on the matched block pairs, we compute:

\textbf{Block-match.}
The ratio of matched block areas to the total area of reference blocks:
\begin{equation}
\text{Block-match} =
\frac{\sum_{(b_r, b_g)\in \mathcal{M}} \mathrm{Area}(b_r \cap b_g)}
{\sum_{b_r \in \mathcal{B}_r} \mathrm{Area}(b_r)}.
\end{equation}

\textbf{Text Similarity.}
Text similarity between matched blocks is measured using the Sørensen--Dice coefficient:
\begin{equation}
\text{Dice}(T_r, T_g) =
\frac{2|T_r \cap T_g|}{|T_r| + |T_g|},
\end{equation}
where $T_r$ and $T_g$ denote character sets extracted by OCR.

\textbf{Color Similarity.}
Perceptual color difference is computed using the CIEDE2000 formula over corresponding blocks.

\textbf{Position Similarity.}
Position similarity measures alignment accuracy based on block center locations:
\begin{equation}
\text{Pos}(b_r, b_g) =
1 - \frac{\|c_r - c_g\|_2}{D},
\end{equation}
where $c_r$ and $c_g$ are block centers and $D$ is the image diagonal length for normalization.

\subsubsection{Component-level Metrics}

\paragraph{Component Segmentation Metrics.}
We evaluate component-level segmentation using one-to-one matching based on Intersection over Union (IoU). For each ground-truth component $g$, the predicted component $p$ with the highest IoU is selected:
\begin{equation}
\text{IoU}(g, p) =
\frac{|g \cap p|}{|g \cup p|}.
\end{equation}

Matched components are counted as True Positives (TP). Unmatched ground-truth and predicted components are counted as False Negatives (FN) and False Positives (FP), respectively. Precision, Recall, and F1 score are computed as:
\begin{equation}
\text{Precision} = \frac{TP}{TP + FP}, \quad
\text{Recall} = \frac{TP}{TP + FN}, \quad
\text{F1} = \frac{2 \cdot \text{Precision} \cdot \text{Recall}}
{\text{Precision} + \text{Recall}}.
\end{equation}

We also report the mean IoU over all matched component pairs as an overall segmentation quality measure.

\paragraph{Component Merging Metrics.}
Component merging is evaluated as a clustering task.

\textbf{Adjusted Rand Index (ARI).}
ARI~\cite{santos2009use} measures the agreement between predicted and ground-truth clusterings while correcting for chance:
\begin{equation}
\text{ARI} =
\frac{\text{RI} - \mathbb{E}[\text{RI}]}
{\max(\text{RI}) - \mathbb{E}[\text{RI}]}.
\end{equation}

\textbf{V-measure.}
V-measure evaluates clustering quality as the harmonic mean of homogeneity $h$ and completeness $c$:
\begin{equation}
\text{V-measure} =
\frac{2hc}{h + c}.
\end{equation}

\subsubsection{Code Metrics}

\paragraph{Tree BLEU}
Tree BLEU~\cite{gui2025webcode2m} evaluates structural similarity between generated and reference HTML by computing the recall of height-1 subtrees in the DOM tree:
\begin{equation}
\text{Tree BLEU} =
\frac{|\mathcal{T}_g \cap \mathcal{T}_r|}
{|\mathcal{T}_r|},
\end{equation}
where $\mathcal{T}_g$ and $\mathcal{T}_r$ denote the sets of subtrees extracted from generated and reference DOMs.

\paragraph{Reuse Rate.}
Let \( B = \{b_1, \dots, b_n\} \) be the set of UI blocks, with \( n = |B| \).
The blocks are partitioned into disjoint groups
\( G = \{g_1, \dots, g_m\} \),
where each group contains blocks that are equivalent under a given criterion. Suppose the set of UI components generated by the model is \(C\), and let \(C_{\ge 2} \subseteq C\) denote the subset of components that are used two or more times. The reuse coverage metric is defined as:


\begin{equation}
\mathrm{Reuse \ Rate}
= \frac{|B| - |G| + |C_{\ge 2}|}{|B|}, 
\end{equation}
The metric yields a normalized value in \([0,1]\), where larger values indicate a higher proportion of blocks participating in code reuse.


\paragraph{Repetitive Ratio via Abstract Component Tree Comparison.}  
The \textit{Repetitive Ratio} quantifies code redundancy by measuring the proportion of code lines corresponding to duplicate UI elements relative to the total lines of code. To identify duplicates, we adopt an \textbf{Abstract Component Tree Comparison} approach:

\begin{enumerate}
    \item \textbf{Component Tree Extraction:} Each UI component is represented as a DOM subtree, preserving tag names and structural hierarchy while ignoring dynamic content such as IDs, inline styles, or textual values.
    \item \textbf{Normalization:} Components are abstracted into tree encodings that capture their structural information.
    \item \textbf{Duplicate Detection:} Two components $T_i$ and $T_j$ are considered duplicates if their tree encodings match exactly. For more flexible matching, we compute the tree edit distance (TED) and define similarity:
    \begin{equation}
    \text{Similarity}(T_i, T_j) = 1 - \frac{\text{TED}(T_i, T_j)}{\max(|T_i|, |T_j|)},
    \end{equation}
    where $|T_i|$ is the number of nodes in the tree. Components with $\text{Similarity}(T_i, T_j) \ge \tau$ (e.g., $\tau = 0.8$) are also considered duplicates.
\end{enumerate}

Once duplicate components are identified, the Repetitive Ratio is computed as:
\begin{equation}
\text{Repetitive Ratio} = \frac{\sum_{k \in \mathcal{D}} L_k}{\sum_{i=1}^{N} L_i} \times 100\%,
\end{equation}
where $\mathcal{D}$ is the set of detected duplicate components, $L_k$ is the number of code lines in component $k$, and $N$ is the total number of components. A lower Repetitive Ratio indicates more compact, modular code with fewer unnecessary duplicates.



\subsection{Human Evaluation Details}
\label{subsec:human_eval}

\textbf{(1) Participant Information and Recruitment:} We recruited five annotators through a local university mailing list. All annotators hold at least a Bachelor's degree in Computer Science or Software Engineering and possess more than two years of professional web development experience. \textbf{(2) Evaluation Procedure:} (i) We randomly sampled 100 webpages from our benchmark, and each annotator evaluated 100 unique pairs comparing direct prompting against one baseline method. (ii) Annotators were required to carefully read the pairwise comparison guidelines (detailed below) before beginning the evaluation. (iii) For each pair, annotators selected one of three options: "Example 1 better", "Example 2 better", or "Tie", following the structured evaluation criteria. To ensure unbiased judgment, annotators were blinded to the method source of each webpage, and pairs were presented in randomized order. \textbf{(3) Inter-rater Reliability:} We computed Krippendorff's alpha to measure inter-annotator agreement, achieving a score of 0.8792. This value exceeds the commonly accepted threshold of 0.8, indicating substantial agreement and confirming the reliability of our human evaluation results.

\begin{tcolorbox}[
    colback=gray!4,
    colframe=black!45,
    title={\centering\textbf{Human Evaluation Instructions}},
    fonttitle=\bfseries,
    boxrule=0.6pt,
    arc=2mm,
    left=1.2mm,
    right=1.2mm,
    top=1.0mm,
    bottom=1.0mm,
    before skip=6pt,
    after skip=6pt,
    skin=enhanced jigsaw
]

\textbf{Task Overview}

In this survey, you will be given a reference webpage's screenshot, as well as two candidate webpages (Example 1 and Example 2) that try to replicate the reference webpage. Your task is to judge which of the two candidates is closer to the reference. Each (Reference, Example 1, Example 2) is presented in a row, where the original boundary of screenshot is marked by black.

\vspace{0.3em}
\textbf{Comparison Guide}

\textbf{Initial Step: Content Check}
\begin{itemize}[leftmargin=*]
\item \textbf{Text Content:} Examine if the text on the candidate webpages matches the reference. Pay special attention to missing or extra content, especially key elements like titles.
\item \textbf{Image Content:} Assess the placement of the blue placeholder blocks (for images).
\item \textbf{Primary Judgment Criterion:} If one example has significant missing or additional content compared to the other, it should be considered less similar to the reference.
\end{itemize}

\textbf{Second Step: Layout Check}
\begin{itemize}[leftmargin=*]
\item \textbf{Element Arrangement:} If the content (text and images) of both examples is similarly good or bad, proceed to evaluate the arrangement of these elements. Check if their organization, order, and hierarchy match the reference.
\item \textbf{Secondary Judgment Criterion:} If differences in layout are observed, the example with the layout most similar to the reference should be rated higher.
\end{itemize}

\textbf{Final Step: Style Check}
\begin{itemize}[leftmargin=*]
\item \textbf{Style Attributes:} Only if Example 1 and Example 2 are comparable in content and layout, examine the style elements like font style, color, and size.
\item \textbf{Tertiary Judgment Criterion:} In cases where content and layout are equally matched, preference should be given to the example with style attributes closer to the reference.
\end{itemize}

\textbf{Overall Judgment}

Based on the criteria in the order of priority (\textbf{Content > Layout > Style}), make an overall judgment on which example (Example 1 or Example 2) is more similar to the reference webpage.

\textbf{Judgment Options}
\begin{enumerate}[leftmargin=*]
\item Select \textbf{"Example 1 better"} if Example 1 is closer to the reference.
\item Select \textbf{"Example 2 better"} if Example 2 is closer to the reference.
\item Opt for \textbf{"Tie"} only if both examples are similarly accurate or equally distant from the reference.
\end{enumerate}

\textbf{Additional Tips}
\begin{enumerate}[leftmargin=*]
\item Use zoom-in for detailed inspection.
\item Focus on major discrepancies in each step before moving to the next.
\item Your judgment should be based on a cumulative assessment of content, layout, and style.
\end{enumerate}

\end{tcolorbox}

\subsection{Prompt Details}
\label{subsec:prompt_detail}

The segmentation prompt is shown in Figure~\ref{fig:seg_prompt}. The UI template code generation prompts and component-based code generation prompts are shown in Figure~\ref{fig:masked_gen_prompt} and Figure~\ref{fig:multi_component_prompt} respectively. The feedback prompt is shown in Figure~\ref{fig:feed_prompt}.
The component-based prompting is shown in Figure~\ref{fig:component_prompt}. 


Figure~\ref{fig:seg_prompt} shows the Semantic Coherent UI Blocks Segmentation Prompt. It guides the MLLM to identify semantically cohesive UI blocks from webpage screenshots. It defines six categories of semantic blocks: (1) block-based components (cards, panels), (2) blocks with associated text, (3) repeated structure groups (product cards, galleries), (4) navigational components (headers, footers), (5) form sections, and (6) content sections (testimonials, pricing tables). The prompt includes strict identification rules (prioritizing semantic meaning, visual hierarchy, natural boundaries) and exclusion rules (avoiding oversized boxes, isolated elements, or splitting unified blocks). The output format requires JSON with bounding boxes and labels.

Figure~\ref{fig:masked_gen_prompt} shows the UI Template Code Generation Prompt, which instructs the model to generate a page layout framework from a masked screenshot where UI block regions are grayed out. 



Figure~\ref{fig:multi_component_prompt} shows the Component-based Code Generation Prompt, which is used for generating reusable components from visually similar UI block screenshots. The prompt requires: (1) creating modular, parameterized components with props for dynamic fields; (2) generating multiple snippet files—one defining the shared reusable component (wrapped in `<component>`), and others showing instance-specific usage (wrapped in `<snippet>`); (3) using loops for repeating content; and (4) replacing images with placeholders. This enables code reuse across similar UI blocks.

Figure~\ref{fig:feed_prompt} show the Element-wise Feedback Prompt, which is designed for fine-grained refinement, this prompt takes: (1) ground truth screenshot with reference annotations, (2) generated UI screenshot with fix markers, (3) current code, and (4) refinement instructions. The model must compare numbered regions, identify add/delete/modify operations, locate corresponding code, and apply fixes. This enables precise, element-level corrections.

Figure~\ref{fig:component_prompt} show the component-based prompting strategy to extract reusable code from webpage screenshots, which is used as a baseline. The prompt instructs the model to: (1) identify component groups based on structural DOM consistency rather than visual appearance, (2) generate standalone Vue components with semantic naming, (3) replace repeating content with loops and dynamic props, and (4) use image placeholders instead of original paths. This approach produces modular, reusable components that can be assembled to reproduce the original webpage layout.



\begin{figure*}[ht]
    \centering
    \includegraphics[width = .95\textwidth]{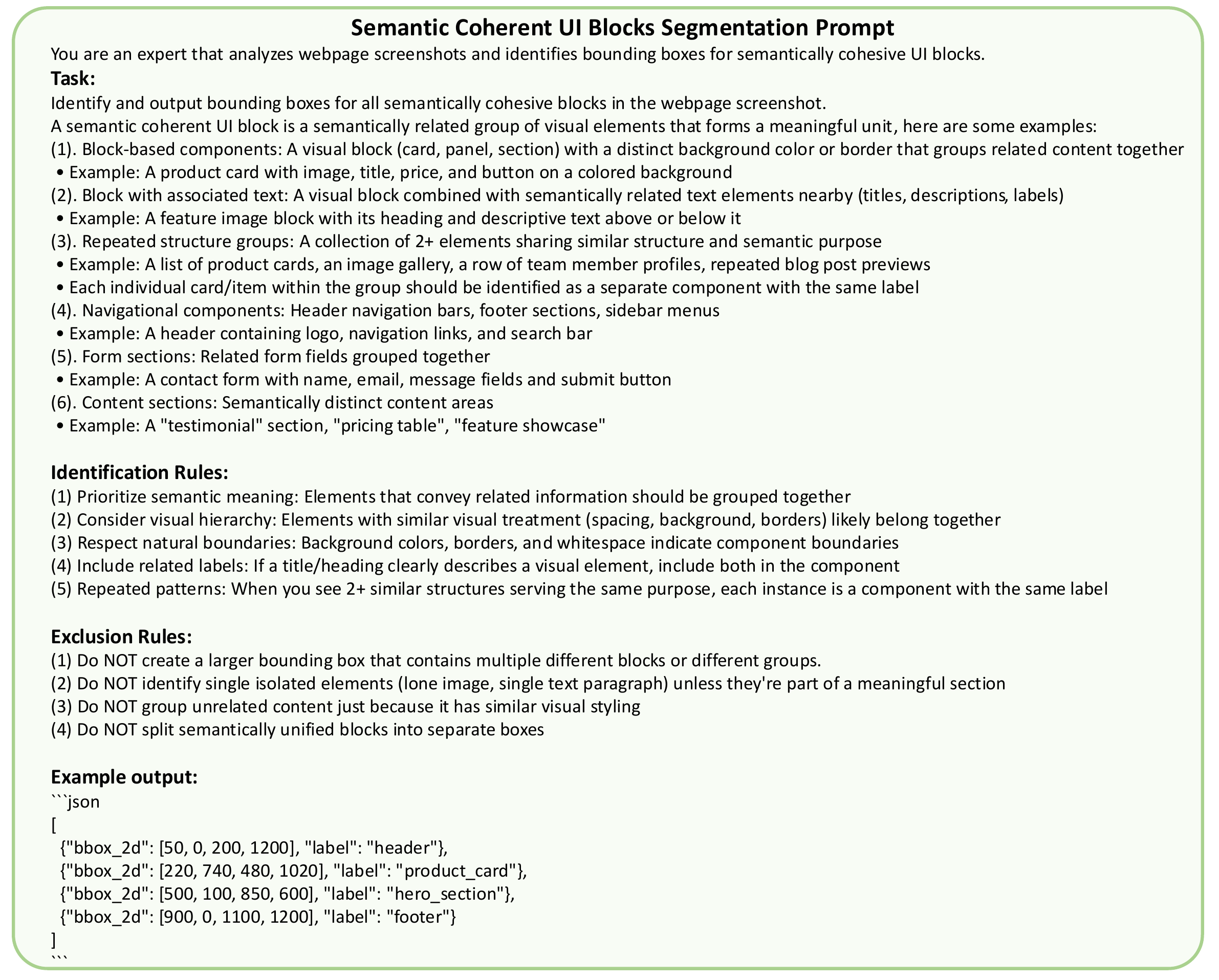}
    \caption{System prompt for detecting semantic coherent UI blocks.}
    \label{fig:seg_prompt}
\end{figure*}

\begin{figure*}[ht]
    \centering
    \includegraphics[width = .8\textwidth]{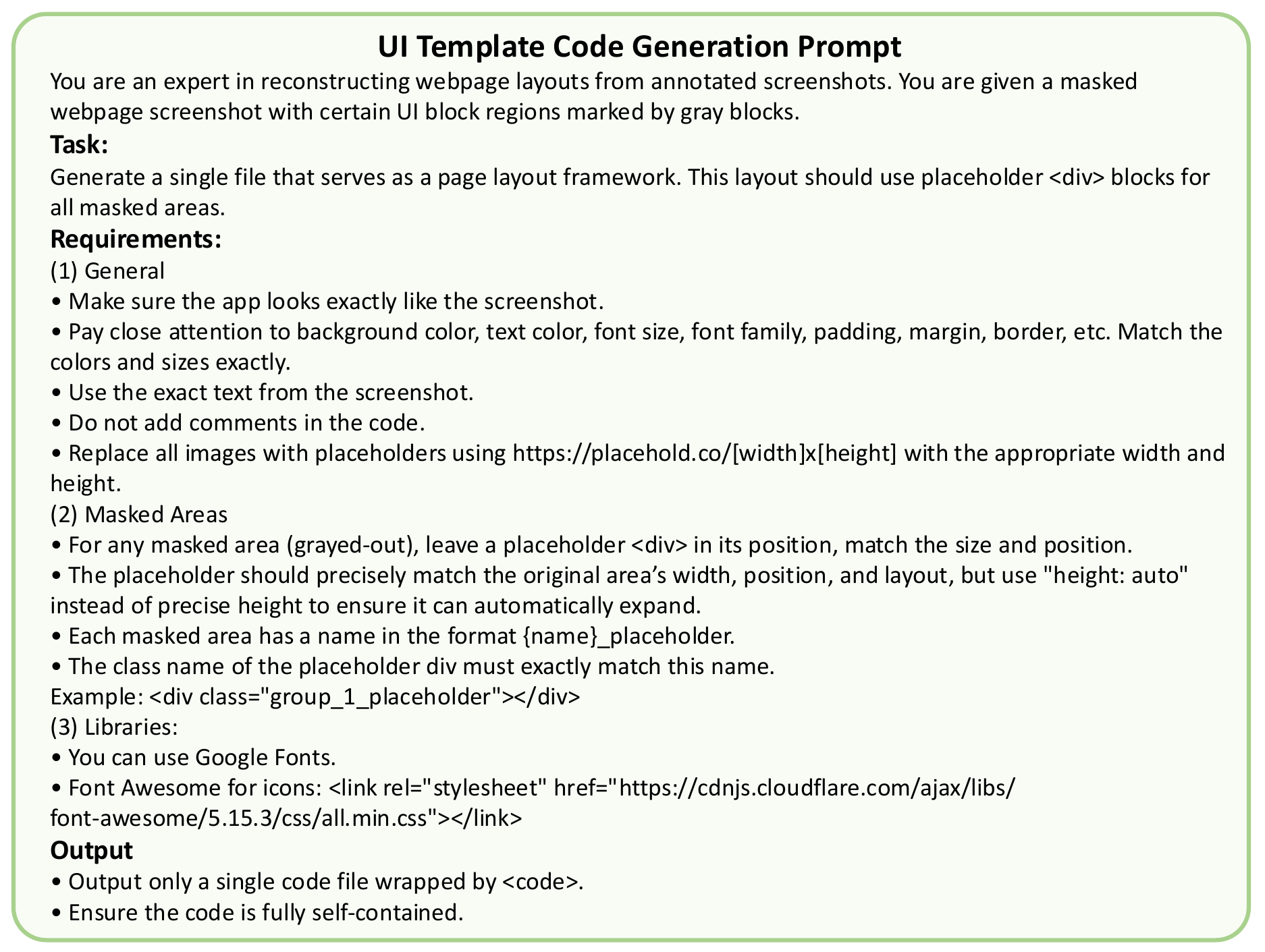}
    \caption{System prompt for generating masked layout framework from partially masked webpage screenshots.}
    \label{fig:masked_gen_prompt}
\end{figure*}

\begin{figure*}[ht]
    \centering
    \includegraphics[width = .8\textwidth]{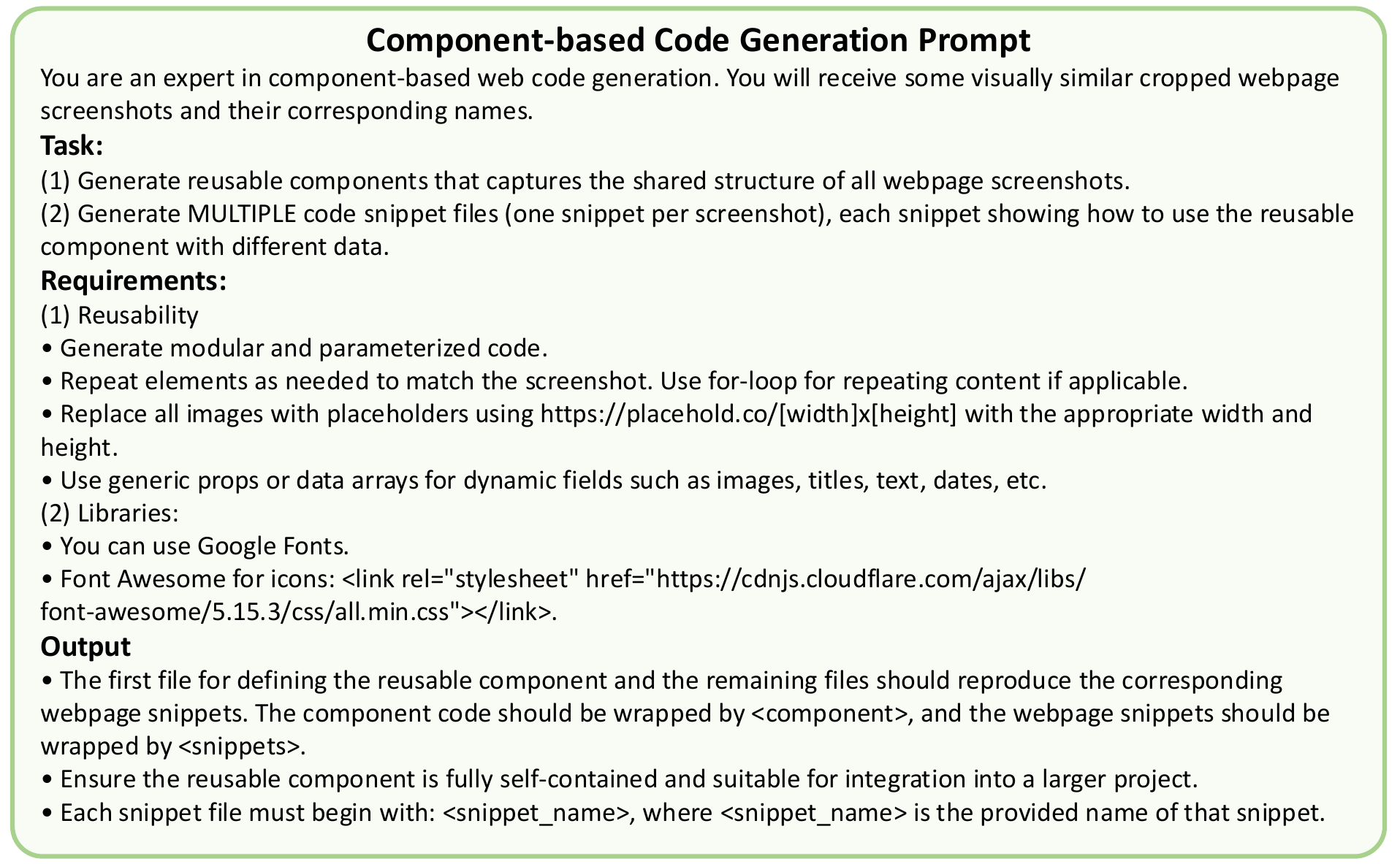}
    \caption{System prompt for component-based code generation from visually similar UI block images.}
    \label{fig:multi_component_prompt}
\end{figure*}

\begin{figure*}[ht]
    \centering
    \includegraphics[width = .8\textwidth]{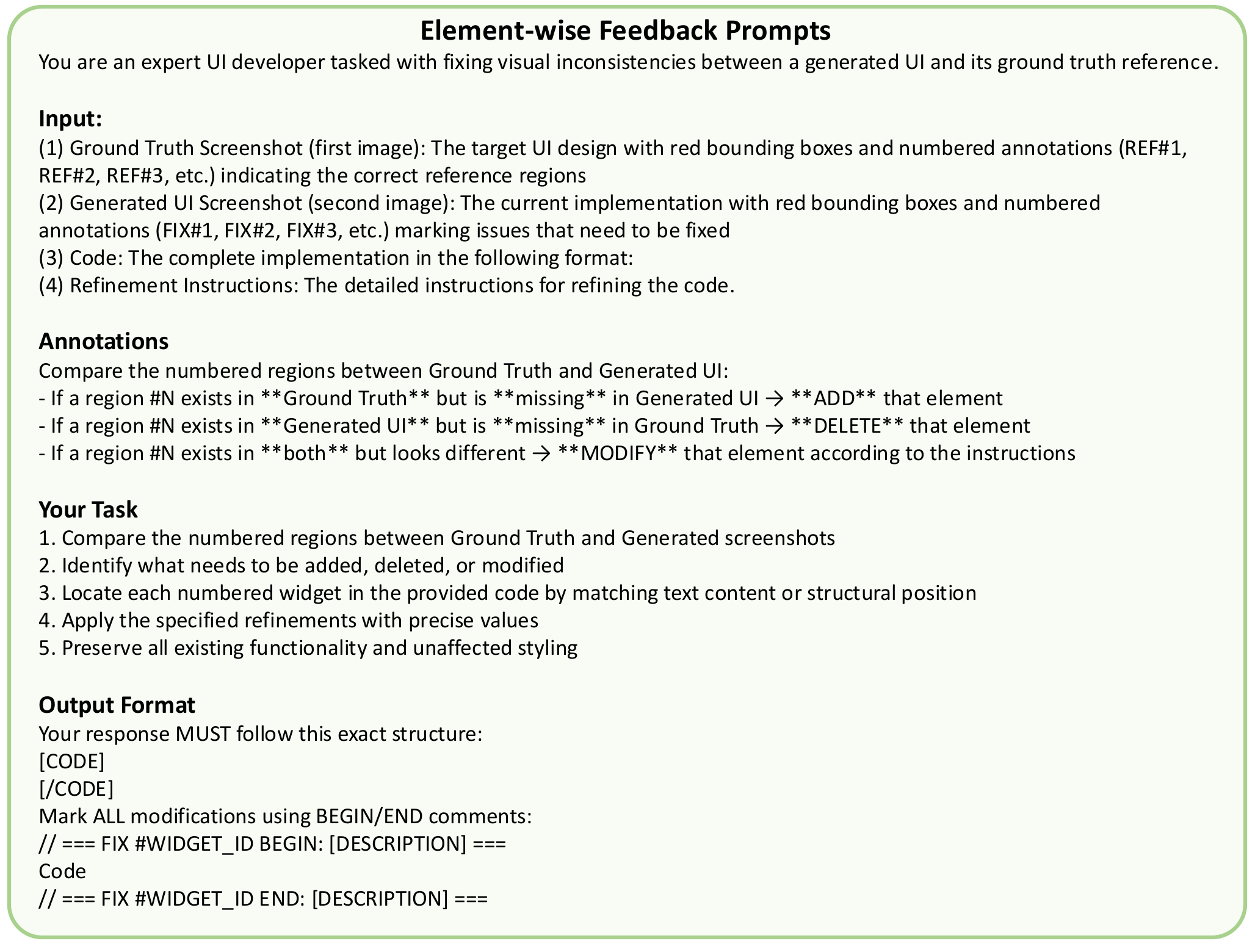}
    \caption{System prompt for element-wise feedback.}
    \label{fig:feed_prompt}
\end{figure*}

\begin{figure*}[ht]
    \centering
    \includegraphics[width = .8\textwidth]{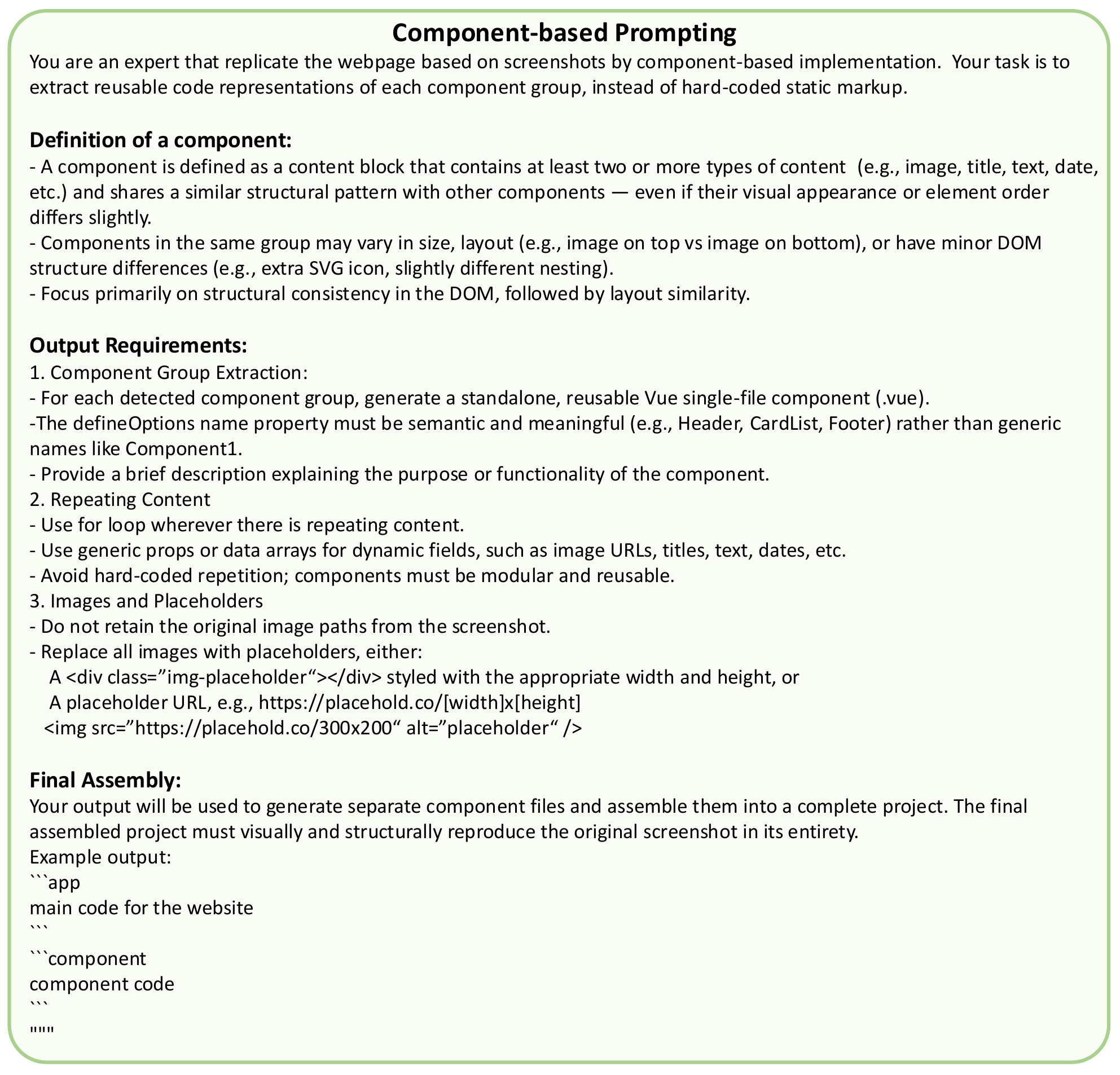}
    \caption{System prompt for end to end component-based code generation (baseline).}
    \label{fig:component_prompt}
\end{figure*}

\subsection{More Cases}
\label{appendix:case_study}


This appendix presents additional qualitative examples that further demonstrate the effectiveness of \scheme \ across segmentation, generation, and feedback refinement.



\paragraph{Case Study One (Figure~\ref{fig:case_study_1}).}
Figure~\ref{fig:case_study_1} presents both segmentation and generation results of different methods on a long and complex multi-section webpage. For segmentation (the first row), \textsc{UICopilot} (a) produces inaccurate bounding boxes with insufficient coverage. \textsc{DCGen} (b) and \textsc{LatCoder} (c) suffer from severe over-segmentation, fragmenting the page into numerous small blocks, while \textsc{LayoutCoder} (d) generates many blank or invalid regions. In contrast, \textsc{ComUICoder} (e) accurately identifies semantically coherent UI blocks, such as headers, banner, artical preview and footers, without unnecessary fragmentation. These segmentation deficiencies further affect downstream generation performance (the second row): \textsc{UICopilot} (a) and \textsc{LatCoder} (c) fail to generate valid webpages due to inaccurate or over-fragmented segmentation; \textsc{DCGen} (b) generates a webpage but exhibits severe layout disorder in the footer and overlapping elements; and \textsc{LayoutCoder} (d) shows layout inconsistencies in the middle section and missing elements in the footer. By contrast, benefiting from high-quality segmentation, \textsc{ComUICoder} (e) successfully generates the complete webpage with a well-structured layout and all elements correctly rendered

\paragraph{Case Study Two (Figure~\ref{fig:case_study_2}).}
The first row of Figure~\ref{fig:case_study_2} demonstrates that \textsc{ComUICoder} (e) achieves semantically coherent segmentation with precisely bounded UI blocks spanning the hero section, feature card, testimonials, and footer navigation. In contrast, competing methods exhibit critical segmentation deficiencies: \textsc{UICopilot} (a) generates only partial page coverage, failing to detect components entirely and the bounding boxes are overlapped with each other; \textsc{DCGen} (b), \textsc{LatCoder} (c) and \textsc{LayoutCoder} (d) produce severe over-segmentation with excessive fragmentation of the feature card, testimonial sections and footer, while introducing some blank blocks.
The second row reveals how segmentation quality directly impacts final output fidelity. \textsc{UICopilot} (a) completely fails to generate the full webpage  due to incomplete segmentation. \textsc{DCGen} (b) produces a webpage with overlapping elements in the feature showcase, text missing in the hero description, and significant layout disorder in the footer navigation. \textsc{LatCoder} (c) exhibits squeezed component proportions, overlap in the footer. \textsc{LayoutCoder} (d) demonstrates wrong alignment in the QA section and layout disorder in the feature section, despite capturing the overall page structure. By contrast, \textsc{ComUICoder} (e) successfully generates a complete and well-structured webpage with accurate element positioning, proper component proportions, and consistent styling throughout all sections, validating the critical role of high-quality segmentation in UI code generation.

\paragraph{Feedback Mechanism Analysis (Figure~\ref{fig:case_feedback}).}
Figure~\ref{fig:case_feedback} illustrates the effectiveness of our feedback mechanism in improving generation quality. The figure compares three versions: (a) the ground truth webpage, (b) \textsc{\scheme} without feedback, and (c) \textsc{\scheme} with feedback enabled. Without feedback (b), the generated webpage exhibits multiple critical defects: wrong alignment in the header section where the three feature cards are mispositioned, missing elements in the pricing table, wrong size of the call-to-action button (``See a Demo''), and missing pagination elements at the bottom of the feature comparison table. These issues significantly degrade the visual fidelity and functional completeness of the generated webpage. In contrast, when the feedback mechanism is activated (c), \textsc{\scheme} successfully corrects all identified defects, producing a webpage that achieves perfect alignment with the ground truth. The header cards are properly positioned, all pricing table elements are present, the call-to-action button displays the correct size and styling, and pagination controls are fully rendered. This case study demonstrates that our feedback mechanism effectively identifies and rectifies generation errors, substantially improving the overall quality and accuracy of the generated UI code.

\begin{figure*}[h]
    \centering
    \includegraphics[width = .8\textwidth]{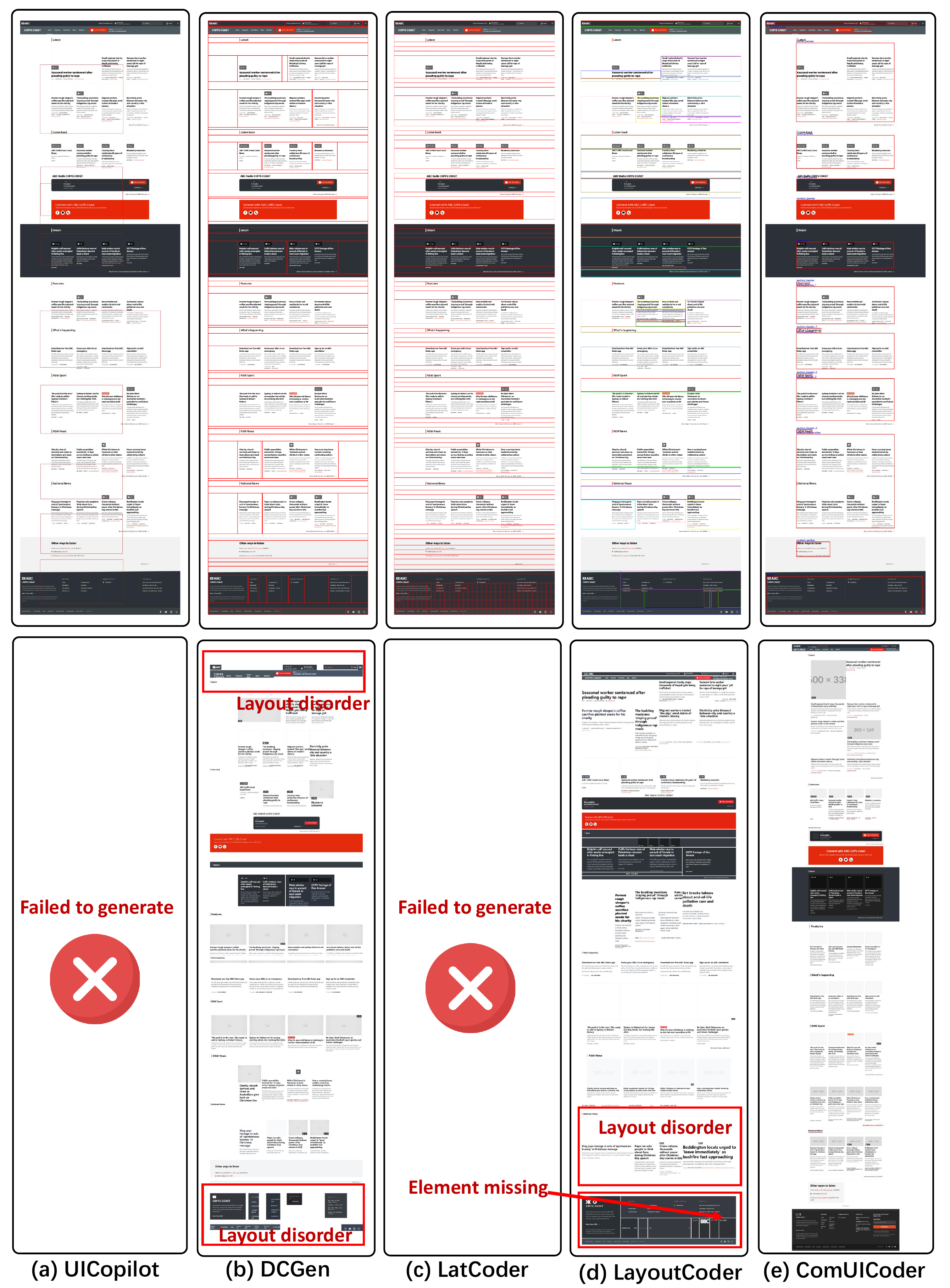}
    \caption{Case one. The first row is the segmentation results and the second row is the generation results. \textsc{ComUICoder} accurately segments semantically coherent UI blocks and generates a complete, well-structured webpage, whereas competing methods suffer from over-fragmentation, missing elements, and layout inconsistencies.
}
    \label{fig:case_study_1}
\end{figure*}

\begin{figure*}[h]
    \centering
    \includegraphics[width = .8\textwidth]{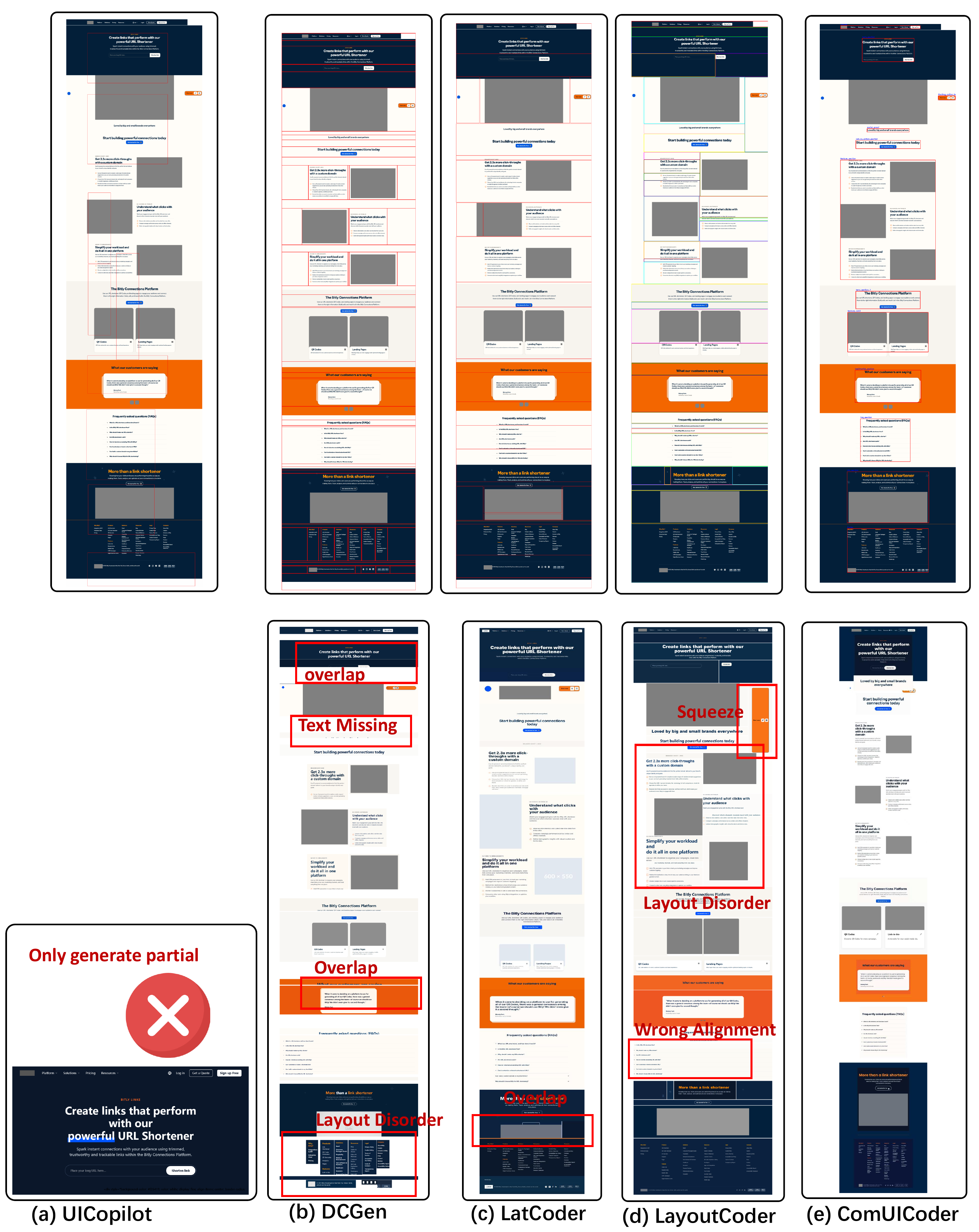}
    \caption{Case two. The first row is the segmentation results and the second row is the generation results. \textsc{ComUICoder} accurately segments semantically coherent UI blocks and generates a complete, well-structured webpage, whereas competing methods suffer from over-fragmentation, missing elements, and layout inconsistencies.}
    \label{fig:case_study_2}
\end{figure*}



\begin{figure*}[h]
    \centering
    \includegraphics[width = .95\textwidth]{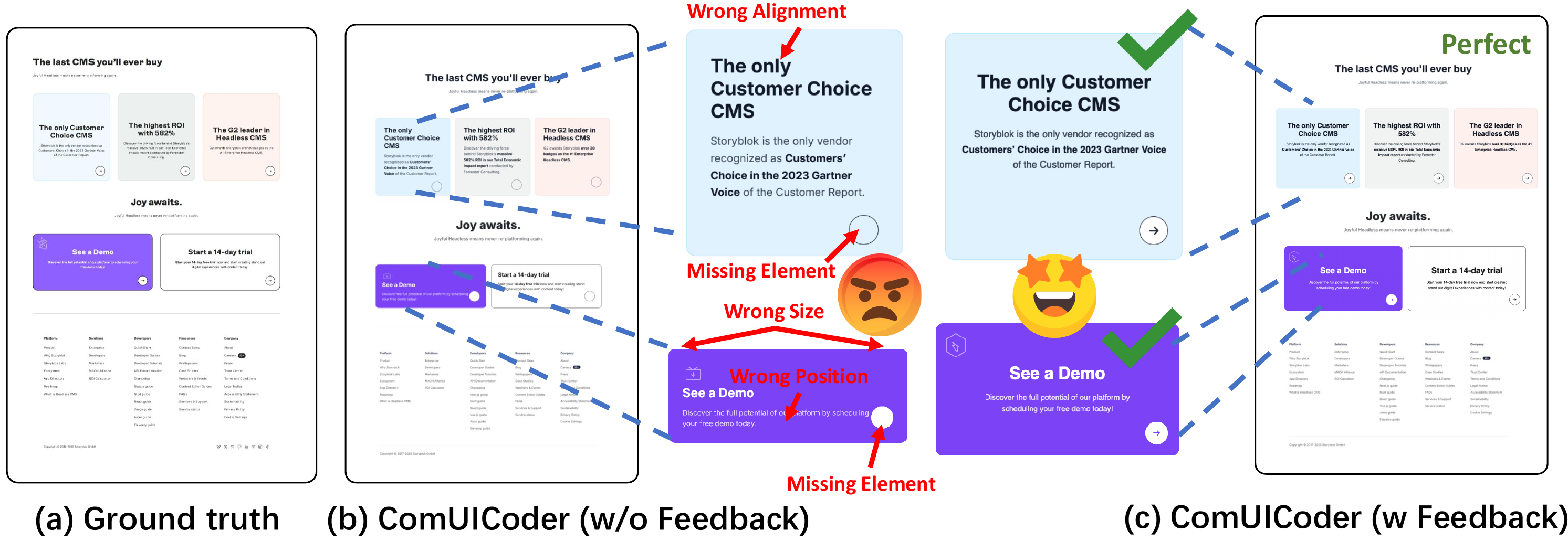}
    \caption{Feedback case. The feedback mechanism further corrects alignment, missing elements, and sizing issues, enhancing visual and functional accuracy of the generated UI code.}
    \label{fig:case_feedback}
\end{figure*}

\end{document}